\documentclass[usenatbib]{mn2e}
\usepackage{graphicx, amssymb, aas_macros, natbib, bm}
\usepackage{ marvosym }
\usepackage{enumitem}

\newcommand{\mstar}{{\rm M}_{\star}}

\newcommand{\msun}{{\rm M}_{\odot}}

\newcommand{\kpc}{{\rm kpc}}
\newcommand{\kms}{{\rm km \, s}^{-1}}

\title[TCB\Lightning: The Mass Dependence of Satellite Quenching]
{Taking Care of Business in a Flash \Lightning:  
Constraining the Timescale for Low-Mass Satellite Quenching with ELVIS
}

\author[Fillingham et al.]
{Sean P. Fillingham,$^1$\thanks{$\!\!$e-mail: sfilling@uci.edu}
Michael C. Cooper,$^1$\thanks{$\!\!$e-mail: cooper@uci.edu}
Coral Wheeler,$^1$ 
\newauthor Shea Garrison-Kimmel,$^1$ 
Michael Boylan-Kolchin,$^2$
James S. Bullock$^1$ \\
$\!\!^1$Center for Cosmology, Department of Physics and Astronomy, 
  4129 Reines Hall, University of California, Irvine, CA 92697 \\ 
$\!\!^2$Department of Astronomy and Joint Space-Science Institute,
University of Maryland, 
College Park, MD 20742-2421}

\begin{document}

\pagerange{\pageref{firstpage}--\pageref{lastpage}} 
\pubyear{2015}

\maketitle

\label{firstpage}
\begin{abstract}
  The vast majority of dwarf satellites orbiting the Milky Way and M31
  are quenched, while comparable galaxies in the field are gas-rich
  and star-forming. Assuming that this dichotomy is driven by
  environmental quenching, we use the ELVIS suite of $N$-body
  simulations to constrain the characteristic timescale upon which
  satellites must quench following infall into the virial volumes of
  their hosts. The high satellite quenched fraction observed in the
  Local Group demands an extremely short quenching timescale
  ($\sim2$~Gyr) for dwarf satellites in the mass range $\mstar \sim
  10^{6}-10^{8}~\msun$. This quenching timescale is significantly
  shorter than that required to explain the quenched fraction of more
  massive satellites ($\sim 8$~Gyr), both in the Local Group and in
  more massive host halos, suggesting a dramatic change in the
  dominant satellite quenching mechanism at $\mstar \lesssim
  10^{8}~\msun$. Combining our work with the results of complementary
  analyses in the literature, we conclude that the suppression of star
  formation in massive satellites ($\mstar \sim 10^{8}-10^{11}~\msun$)
  is broadly consistent with being driven by starvation, such that the
  satellite quenching timescale corresponds to the cold gas depletion
  time. Below a critical stellar mass scale of $\sim 10^{8}~\msun$,
  however, the required quenching times are much shorter than the
  expected cold gas depletion times. Instead, quenching must act on a
  timescale comparable to the dynamical time of the host halo.  We
  posit that ram-pressure stripping can naturally explain this
  behavior, with the critical mass (of $\mstar \sim 10^{8}~\msun$)
  corresponding to halos with gravitational restoring forces that are
  too weak to overcome the drag force encountered when moving through
  an extended, hot circumgalactic medium.
\end{abstract}


\section{Introduction}
\label{sec:intro} 

Foremost among the results of galaxy surveys over the last decade has
been the realization that the galaxy population at $z \lesssim 2$ is
bimodal in nature \citep[e.g.][]{strateva01, baldry04, bell04,
  cooper06}. That is, galaxies both locally and out to intermediate
redshift are effectively described as one of two distinct types: red,
early-type galaxies lacking significant star formation and blue,
late-type galaxies with active star formation. In color-magnitude
space, the red galaxies populate a tight relation (often called the
red sequence), while the distribution of blue galaxies is more
scattered (sometimes referred to as the blue cloud).
While the red and blue populations comprise approximately equal
portions of the cosmic stellar mass budget at $z \sim 1$, galaxies on
the red sequence dominate today, following a growth in stellar mass
within the red population of roughly a factor of $2$ over the past $7$
Gyr \citep{bell04, bundy06, faber07, brown07}.
Despite uncertainty regarding the particular physical process(es) at play,
the suppression (or quenching) of star formation in blue galaxies,
thereby making them red, is one of the principal drivers of this
dramatic growth in the number density of quiescent systems at late
cosmic time.

At both low and intermediate redshift, the local environment of a
galaxy is known to be well-correlated with the suppression of star
formation, such that passive or quiescent galaxies preferentially live
in higher-density environments \citep{balogh04, kauffmann04, blanton05,
  cooper06, cooper07, cooper10b}.
While the stellar mass or surface mass density of a galaxy may be more
closely connected to quenching for massive systems \citep{peng10,
  cheung12, woo13}, recent work has shown that environment is likely
the dominant driver of quenching at the lowest mass scales
\citep[$\mstar < 10^9~\msun$,][]{geha12}.  For example, studies
comparing satellite galaxies to isolated field systems of similar
stellar mass in the local Universe find that satellites tend to
exhibit lower star formation rates, more bulge-dominated morphologies,
as well as older and more metal-rich stellar populations
\citep{baldry06, vdb08, cooper10a, pasquali10, tollerud11,phillips14}. This
observed suppression of star formation in satellite galaxies is
commonly referred to as ``environmental quenching''.
At present, galaxy formation models fail to reproduce this
environment-dependent effect in detail, with both semi-analytic and
hydrodynamical models overpredicting the number of quenched satellites
at low masses \citep{kimm09, weinmann11, weinmann12, hirschmann14,
  wang14}.

Understanding the physics of environmental quenching is complicated by
the wide range of physical processes that could be responsible for
suppressing star formation in low-mass satellites. For example,
various quenching mechanisms preferentially operate on satellite
galaxies within overdense environments, including [\emph{i}]
``starvation'' or ``strangulation'' -- whereby gas accretion onto a
satellite galaxy is halted following infall, thus eventually
eliminating the fuel for star formation \citep{larson80, kawata08},
[\emph{ii}] ``harassment'' -- by which close encounters between
densely packed cluster or group members strip gas from around the
interacting galaxies \citep{moore96}, and [\emph{iii}] ``ram-pressure
stripping'' -- where the cold dense gas at the center of a satellite
(or the hot halo surrounding it) is 
removed from the galaxy as a result of a high-speed interaction with
the hot gas halo of the host \citep{gunn72, bekki09}.

To further complicate matters, each of these processes may be
important in different host halos and at different satellite mass
scales.
In an attempt to differentiate between these various physical
mechanisms, we aim to measure the timescale upon which satellite
quenching occurs across a broad range of satellite stellar
mass. Throughout this work, we define this quenching timescale
($\tau_{\rm quench}$) relative to the time at which a satellite was
accreted onto its current host system (i.e.~relative to the infall
time of the satellite). While quenching via starvation will proceed
according to the available cold gas reservoir for star formation
within an accreted satellite, ram-pressure stripping will largely act
on a timescale set by the density distribution of the host's hot halo
and infall velocity of the satellite. As such, constraining the
timescale for satellite quenching may therefore serve as a critical
step towards differentiating between various physical mechanisms.

By comparing satellite samples at low and intermediate redshift to
galaxy and dark matter-only simulations, several recent analyses of
satellites at high stellar mass ($> 5 \times 10^{9}~\msun$) point
towards relatively long quenching timescales of several Gyr or more
\citep{wetzel13, delucia12}. 
Remarkably, when pushing this analysis to lower masses ($\sim 10^{8.5}
- 10^{9.5}~\msun$), \citet{wheeler14} find that quenching remains
surprisingly inefficient, such that the average quenching timescale
increases to $> 7$~Gyr. Here, we extend these previous analyses of
quenching timescales to yet lower masses by studying the satellite
galaxies of the Local Group. Through comparison to simulations, we aim
to establish a coherent picture of the timescale upon which satellite
quenching acts, spanning more than five orders of magnitude in
satellite mass. In \S\ref{sec:LG}, we detail the population of Local
Group satellites along with an estimation of the observed quenched
fraction. In \S\ref{sec:ELVIS} and \S\ref{sec:models}, we describe the
relevant $N$-body simulations and physically-motivated quenching
models, respectively, detailing the relevant constraints on the
derived quenching timescale at low masses. Finally, in
\S\ref{sec:discussion} and \S\ref{sec:summary}, we discuss the
implications of our analysis, including a ``back of the envelope''
explanation of the underlying physics operating on low-mass satellites
and a summary of our main results.


\section{Local Group Satellites}
\label{sec:LG}

\begin{figure}
 \centering
 \hspace*{-0.05in}
 \includegraphics[width=3.35in]{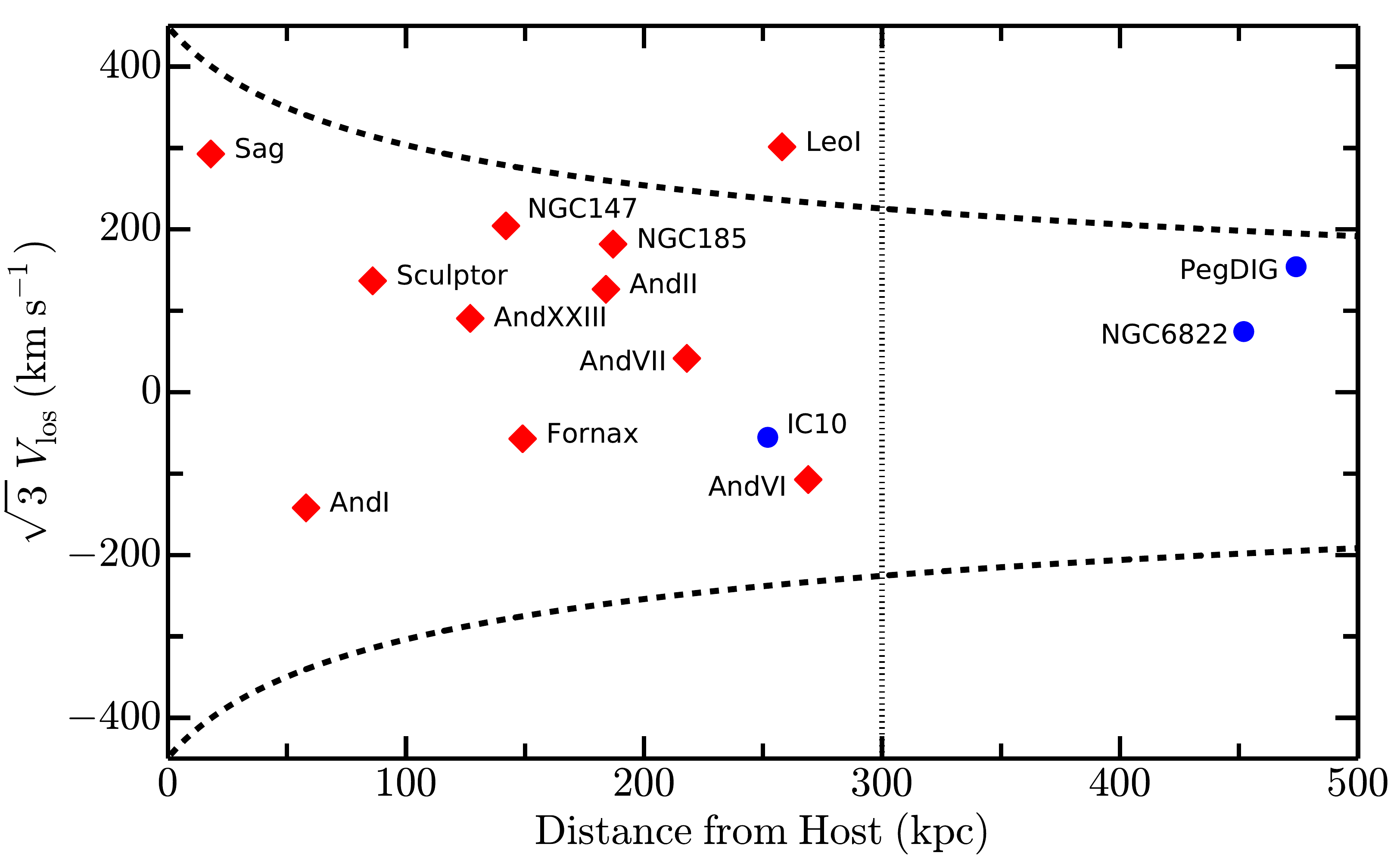}
 \caption{Host-centric velocity versus distance for all known Local
   Group galaxies in the stellar mass range of
   $10^{6}-10^{8}~\msun$. To account for unknown tangential
   velocities, the observed line-of-sight velocities have been multiplied by a
   factor of $\sqrt{3}$. The dotted vertical line at $300$~kpc
   corresponds to the expected virial radius of the Milky Way and M31
   and represents the radial limit for our satellite selection. The
   dashed lines show the escape velocity for an NFW halo with a
   concentration ($c$) of $8$ and a total mass of $2\times 10^{12}\
   \rm M_{\sun}$. Satellites are color-coded as star-forming (blue
   circles) or quenched (red diamonds) according to their observed
   H{\scriptsize I} gas fraction (see
   Fig.~\ref{fig:h1}). }
 \label{fig:vlos}
\end{figure}

In assembling our sample of Milky Way and M31 satellites, we utilize
the compilation of local dwarf galaxy properties presented by
\citet{mcconnachie12}. To minimize the potential impact of
incompleteness at low masses, we limit our analysis to those dwarfs
with measured stellar masses in the range $10^{6}-10^{8}~\msun$;
however, as discussed in more detail below, including systems at lower
masses has very little impact on our qualitative or quantitative
results. From this population of local dwarfs, satellite systems are
selected to be within $300$~kpc of the Milky Way or M31 and bound with
respect to their host --- i.e.~$\sqrt{3} \cdot V_{los} < V_{\rm
  esc}(r)$, where $V_{\rm esc}(r)$ is the escape velocity at a radial
distance of $r$ for an NFW halo with a concentration ($c$) of $8$ and
a total halo mass of $2\times 10^{12}\ \rm M_{\sun}$.\footnote{Varying
  the adopted host halo mass and concentration (and thus our satellite
  selection limits) does not have a significant effect on our final
  sample of Local Group satellites.} To account for potential
tangential motion, we multiply all observed line-of-sight velocities
from \citet{mcconnachie12} by a factor of $\sqrt{3}$ in evaluating
whether each system is bound. Within the stellar mass range of
$10^{6}~\msun$ to $10^{8}~\msun$, Figure~\ref{fig:vlos} shows the
position and velocity of all Local Group dwarfs relative to their host
(or nearest massive neighbor in the case of field systems such as the
Pegasus dIrr).

We identify $12$ satellite systems meeting our selection criteria (see
Table~1), with the final set of satellites and
corresponding quenched fraction relatively independent of the adopted
selection criteria. For instance, field dwarfs such as NGC 6822, Leo
T, and Phoenix, which are nearly in our stellar mass limits, are
located significantly beyond our radial selection limit
(i.e.~$>~\!\!\!~400$~kpc from the Milky Way). Moreover, while the
observed line-of-sight velocity of Leo I suggests that it may be
unbound (see Fig.~\ref{fig:vlos}), recent proper motion observations
conclude that the system is very likely bound to the Milky Way even at
a distance of more than 250~kpc \citep{sohn13, bk13}. Finally, given
the high metallicity and lack of overdensity in RR Lyrae stars, we
categorize Canis Major as an overdensity in the Milky Way and not a
satellite \citep[][but see also \citealt{md05}]{momany04, butler07,
  mateu09}.

\begin{figure}
 \centering
 \hspace*{-0.08in}
 \includegraphics[width=3.35in]{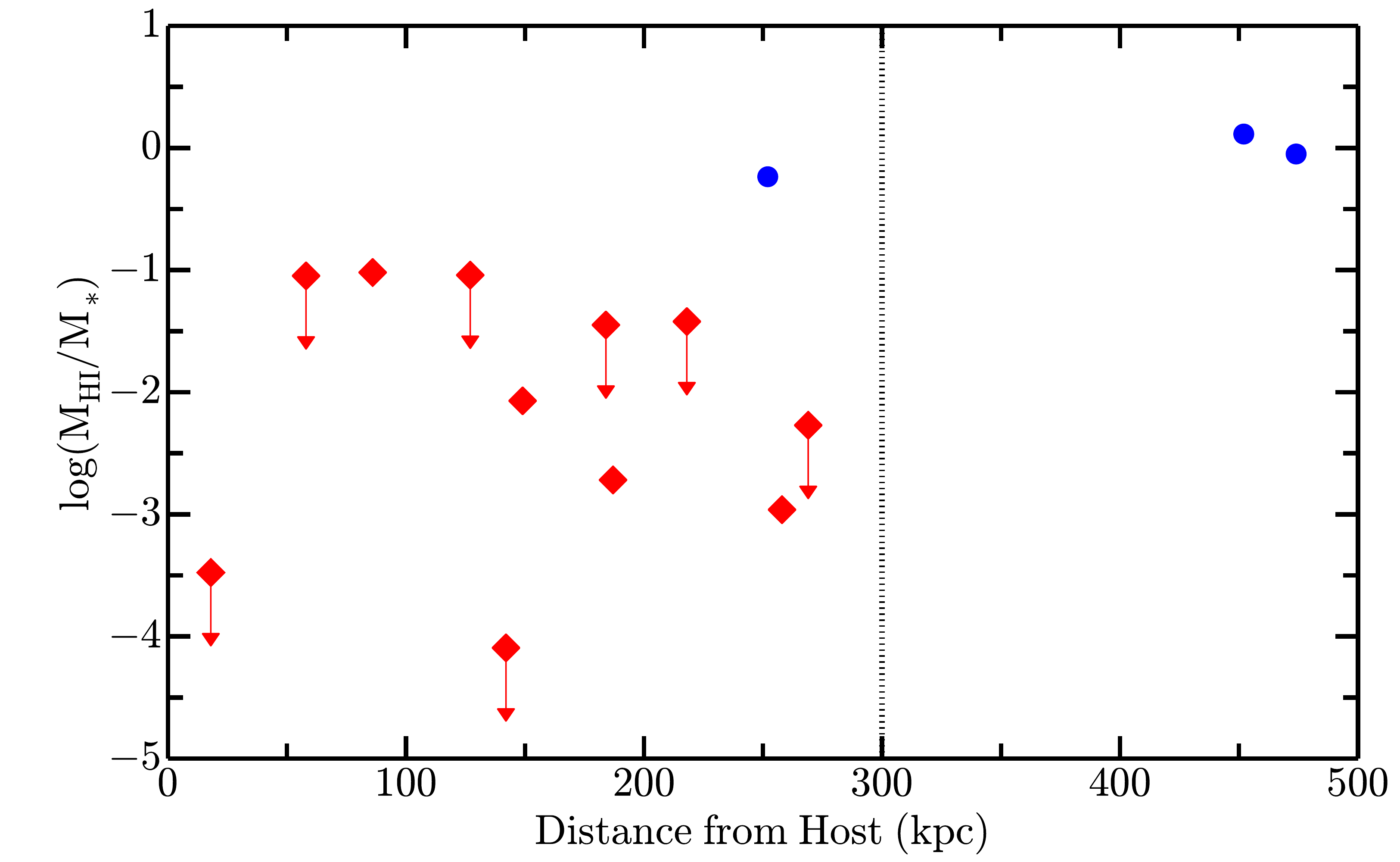}
 \caption{Observed H{\scriptsize I} gas fraction as a function of
   distance from the nearest host galaxy for local dwarfs included in
   our mass range of $10^6-10^{8}~\msun$. Systems with an atomic gas
   fraction less than $10\%$ are classified as quenched (red
   diamonds), with more gas-rich dwarfs identified as star-forming
   (blue circles). At these low masses, roughly $90\%$ of satellites
   are quenched in the Local Group.}
 \label{fig:h1}
\end{figure}

To determine whether a given satellite is quenched, we use the
observed H{\scriptsize I} gas fraction as a proxy for the current star
formation rate (SFR), defining dwarfs with atomic gas fractions less
than $10\%$ to be quenched (see Fig.~\ref{fig:h1}). This criterion
leads to only one satellite in our sample, IC 10, categorized as
currently star-forming and yields a quenched fraction of $>90\%$ for
satellites bound to the Milky Way and M31 in the selected mass
range. 
Alternative definitions of quenched versus star-forming -- e.g.~using
star-formation histories derived from photometric observations of
spatially-resolved stellar populations \citep{weisz14a, weisz15} or
star-formation rates inferred from H$\alpha$ emission \citep{kaisin13,
  karachentsev13} yield similar results.
Moreover, our measurement of the satellite quenched fraction is not
strongly dependent on the adopted mass range, as the quenched fraction
is largely unchanged if we extend the lower mass limit down to
$3~\times~10^{5}~\msun$ so as to account for potential uncertainties
or scatter in the stellar mass measurements.
Over this expanded stellar mass range, the sample would then also
include Leo II, And XXI, And XXV, and LGS~3. While H{\scriptsize I}
measurements suggest that LGS~3 is likely star-forming, a variety of
observations, including star formation rates inferred from H$\alpha$
narrow-band imaging, indicate that Leo II, And XXI, and And XXV are
quiescent \citep{grcevich09, kaisin12, kaisin13, spekkens14, weisz14a,
  weisz14b}. When including these four lower-mass systems in our
sample, the resulting quenched fraction remains quite
high 
($\sim87\%$); furthermore, pushing to yet lower masses will similarly
yield little variation in the quenched fraction, as the lowest-mass
satellites in the Local Group are universally quenched
\citep[e.g.][]{okamoto08, sand09, sand10, weisz14a, brown14}.

While our census of satellites in the Local Group is likely incomplete
at low masses due to obscuration by or confusion with the disk of both
the Milky Way and M31, the focus of our analysis is the relative
fraction of star-forming versus quenched systems. Assuming that
neither quenched nor star-forming systems preferentially reside in or
behind the disk, the quenched fraction should be independent of any
such incompleteness. From an analysis of the Palomar and UK Schmidt
photographic plates, which cover more than $20,000$~square~degrees of
sky at high Galactic latitude ($|b| > 20^{\circ}$), \citet{irwin94}
finds that the known satellite population for the Milky Way is
complete down to below our adopted mass limit
($\gtrsim~3~\times~10^{5}~\msun$). This conclusion is supported by
deeper imaging data from the Sloan Digital Sky Survey
\citep[SDSS,][]{york00}, Pan-STARRS \citep{kaiser10}, and Dark Energy
Survey \citep[DES,][]{DES14}, which cover a significant portion of the
northern and southern sky and are complete to beyond the host virial
radius (i.e.~$>~\!~300$~kpc) at these stellar masses
\citep{tollerud08, koposov08, laevens14, koposov15, DES15}.

\begin{table*}
\centering 
\begin{tabular}{l c c c c c c } 
  \hline\hline 
  Galaxy Name & Morphology & $\mstar~(10^{6}~\msun)$ &
  ${\rm M}_{\rm H{\textsc{\tiny I}}}~(10^{6}~\msun)$ & $D_{\rm host}$ (kpc) &
  $V_{\rm los}~({\rm km}~{\rm s}^{-1})$ & Quenched \\
  & (1) & (2) & (3) & (4) & (5) & (6) \\
  \hline\hline
  Milky Way &  & & & & & \\
  \hline
Sagittarius & dSph & 21 & $< 0.007$ & 18 & 169 & Yes \\
Sculptor & dSph & 2.3 & 0.22 & 86 & 79 & Yes \\
Fornax & dSph & 20 & 0.17 & 149 & -33 & Yes \\
Leo I & dSph & 5.5 & $< 0.006$ & 258 & 174 & Yes \\
\hline
M31 &   & & & & & \\
\hline
And I & dSph & 3.9 & $< 0.35$ & 58 & -82 & Yes \\
And XXIII & dSph & 1.1 & $<0.1$ & 127 & 52 & Yes \\
NGC 147 & dE/dSph & 62 & $< 0.003$ & 142 & 118 & Yes \\
And II & dSph & 7.6 & $< 0.27$ & 184 & 73 & Yes \\
NGC 185 & dE/dSph & 68 & 0.13 & 187 & 105 & Yes \\
And VII & dSph & 9.5 & $< 0.36$ & 218 & 24 & Yes \\
IC 10 & dIrr & 86 & 50 & 252 & -32 & No \\
And VI & dSph & 2.8 & $< 0.015$ & 269 & -62 & Yes \\
\hline 
\hline
\end{tabular} 
\label{table:dwarfs} 
\caption{Observed properties of satellite galaxies in the Local Group
  with measured stellar masses in the range $10^{6}-10^{8}~\msun$ and
  located within $300~\kpc$ of either the Milky Way or M31 (separated
  according to host system):  
  (1) morphological classification and (2)
  stellar mass from \citet{mcconnachie12}; (3)
  atomic gas mass measurements from \citet{mcconnachie12} and
  \citet{grcevich09}, plus upper limits from the literature
  \citep{burton99, grcevich09, huang12a, giovanelli13};
  (4) distance from the nearest host as given by \citet{mcconnachie12};   
  (5) line-of-sight velocity with respect to the host from
  \citet{mcconnachie12}, with the exception of And~XXIII 
  \citep{kirby14}; (6) identification as quenched 
  versus star-forming, according to observed atomic gas fraction
  (see~Fig.~\ref{fig:h1}). } 
\end{table*}

For M31, our knowledge of satellites is largely based on imaging
surveys such as the Pan Andromeda Archeological Survey
\citep[PAndAS,][see also
\citealt{ferguson02,ibata07,mcconnachie08}]{mcconnachie09}. As shown
by \citet{brasseur11}, PAndAS is complete down to our mass limit
within a survey footprint that covers a radial distance of roughly
$150$~kpc surrounding M31. While recent analysis of imaging data,
including that from Pan-STARRS, has likely uncovered several new
satellites of M31 at radial distances of $\gtrsim 150$~kpc \citep[And
XXX-XXXIV,][]{conn12, martin13a, martin13b, martin14}, we exclude
these systems from our study given that distances (relative to M31)
are poorly constrained and potentially greater than $300$~kpc. Early
analysis of the stellar populations in Andromeda XXXI and XXXII,
however, do suggest passive populations devoid of young stars
\citep{martin13a}. As likely quenched satellites, we note that
inclusion of And XXX-XXXIV would have little impact on the observed
quenched fraction in the Local Group and thus on the results of our
analysis. Moreover, while the M31 satellite population may be
incomplete at radial distances of $150 - 300$~kpc from M31, this
should have minimal impact on the observed quenched fraction in the
Local Group (even assuming that star-forming satellites are possibly
biased towards large radial distances, as our results would imply). In
\S\ref{sec:models}, we discuss this potential selection effect in more
detail.


\section{Simulations}
\label{sec:ELVIS}

To model the evolution of satellite galaxies in the Local Group, we
utilize the Exploring the Local Volume In Simulations (ELVIS) suite of
$48$ high-resolution, dissipationless simulations of Milky Way-like
halos \citep{gk14}. The suite includes $24$ isolated halos as well as
$12$ mass-matched Local Group-like pairs, simulated within
high-resolution uncontaminated volumes spanning $2$-$5$~Mpc in size
using a particle mass of $1.9 \times 10^{5}~\msun$ and a
Plummer-equivalent force softening of $\epsilon = 141$~physical
parsecs. Within the high-resolution volumes, the halo catalogs are
complete down to ${\rm M}_{\rm halo} > 2 \times 10^{7}~\msun$, $V_{\rm
  max} > 8~\kms$, ${\rm M}_{\rm peak} > 6 \times 10^{7}~\msun$, and
$V_{\rm peak} > 12~\kms$ --- thus more than sufficient to track the
evolution of halos hosting Local Group dwarfs with stellar masses of
$> 10^{6}~\msun$. ELVIS adopts a cosmological model based on
{Wilkinson Microwave Anisotropy Probe 7} \citep{komatsu11}, with the
following $\Lambda$CDM parameters: $\sigma_{8}=0.801$,
$\Omega_{m}=0.266$, $\Omega_{\Lambda}=0.734$, $n_{s}=0.963$, and
$h=0.71$.

As hosts for the Local Group dwarfs in the stellar mass range of
$10^{6}-10^{8}~\msun$, we select halos in the ELVIS simulations with
masses of ${\rm M}_{\rm peak} = 5\times10^{9}-6\times10^{10}~\msun$,
following the abundance matching prescription of
\citet{gk14}. Abundance matching, a common technique for populating
simulated dark matter distributions with galaxies, assumes a
one-to-one relation between a galaxy's stellar mass and the mass of
its parent dark matter (sub)halo \citep{behroozi13, moster13}. While
this simple empirical approach to modeling galaxy formation yields
great success for massive galaxies, matching a wide range of
clustering statistics as a function of cosmic time
\citep[e.g.][]{berrier06, conroy06}, it potentially breaks down at low
masses, where the most massive subhalos of Milky Way-like simulations
are inconsistent with the observed dynamics of nearby dwarf galaxies
\citep{bk11, bk12, gk14b, kirby14, tollerud14}.

While uncertainties remain in the potential efficacy or robustness of
abundance matching prescriptions at low masses, our results are
largely unaffected by changes in the assumed stellar mass-halo mass
relation. As our fiducial model, we adopt the abundance matching
relation of \citet{gk14}, which in our stellar mass range ($10^{6} -
10^{8}~\msun$) yields $9$ subhalos as candidate satellite hosts within
a typical ELVIS parent halo (after removing stripped or disrupted
subhalos). When applying the shallower stellar mass-halo mass relation
from \citet{behroozi13}, the number of potential subhalo hosts
increases accordingly, but the quantitative and qualitative results
with regard to satellite quenching remain unchanged. To account for
potential increased scatter in the abundance matching relation at low
masses, we also mimic scatter in the \citet{gk14} relation, by
broadening the range of halo masses associated with the observed
stellar mass range, randomly selecting $9$ subhalos within $2 \times
10^{9} < {\rm M}_{\rm peak}/\msun < 2 \times 10^{11}$ as satellite
hosts. Again, while this biases the subhalo population to lower
masses, our results with regard to satellite quenching remain
unaffected.

Independent of the particular abundance matching prescription, we
account for satellites that have been tidally-stripped or disrupted
--- and thus not included in our observational sample --- by
restricting our analysis to subhalos for which $V_{\rm max}/V_{\rm
  peak} > 0.3$, where $V_{\rm max}$ is the current ($z=0$) maximum
circular velocity for the halo and $V_{\rm peak}$ is the value of
$V_{\rm max}$ when the halo mass is at its maximum. While a halo's
maximum halo mass generally occurs prior to infall (typically at $>
1.5~R_{\rm vir}$), $V_{\rm peak}$ is a good approximation for $V_{\rm
  infall}$ since a halo's mass, in general, does not change
appreciably prior to infall \citep{behroozi14}. By restricting our
analysis to present-day halos with $V_{\rm max}/V_{\rm peak} > 0.3$,
we eliminate all subhalos (i.e.~satellites) that have lost greater
than $90\%$ of their mass since infall, thereby selecting those
subhalos that are within the desired mass range and are highly likely
to currently host dwarf galaxies (i.e.~not disrupted or significantly
stripped). Finally, we note that our results are not strongly
dependent upon the adopted $V_{\rm max}/V_{\rm peak}$ cut; applying a
more conservative selection criterion (e.g.~$V_{\rm max}/V_{\rm peak}
> 0.5$) does not significantly change our results.

The ELVIS merger trees include $75$ snapshots from $z = 125$ to $z =
0$, providing time resolution of roughly $200$~Myr. As discussed in
more detail in \S\ref{sec:models}, we are interested in using ELVIS to
explore models where satellite quenching occurs at a fixed physical
radius from the host (e.g.~within $50$~kpc radially). The crossing
time on such scales ($\sim100$~kpc), however, is typically greater
than the median time resolution of the snapshots, such that subhalos
would potentially pass into and back out of the quenching radius
between two successive snapshots. To more precisely determine subhalo
orbits within ELVIS, we map the spatial position of each subhalo
(relative to its host) at $20$~Myr intervals by spline interpolating
the position of each subhalo and corresponding host halo across each
of the $75$ snapshots. This interpolation also allows us to determine
the time at which each subhalo was accreted onto its host halo
(i.e.~the infall time or $t_{\rm infall}$) at greater precision. Due
to the linear spacing of the ELVIS snapshots with scale factor
(i.e.~non-linear spacing with lookback time), the difference in the
measured infall time between our interpolated data and that inferred
from the standard ELVIS merger trees depends upon the lookback time at
which a subhalo is accreted. For systems with $t_{\rm infall} <
3$~Gyr, which place the strongest constraints on the satellite
quenching timescale, the difference in $t_{\rm infall}$ is relatively
modest, with a median offset of less than $0.5$~Gyr for our fiducial
sample of subhalos.


\section{Quenching Models}
\label{sec:models}

As highlighted in \S\ref{sec:intro}, observations of low-mass galaxies
in the local Universe suggest that all isolated (or ``field'') systems
at $\lesssim 10^{9}~\msun$ are star-forming. Thus, all quenching at
these low masses is likely driven by environmental effects associated
with dwarf galaxies being accreted onto more massive halos, such as
that of the Milky Way or M31. Two likely quenching mechanisms
operating on satellite systems are starvation, where future gas
accretion onto the satellite is halted following infall into a more
massive host halo, and ram-pressure stripping, where the cold gas in a
satellite galaxy is violently removed via interaction with the hot gas
halo of the host. To mimic these respective processes, we implement
two quenching models in the ELVIS simulations: (\emph{i})~a
``starvation'' model in which quenching depends only on time since
infall into the host halo and (\emph{ii})~a ``stripping'' model in
which quenching occurs once a satellite reaches a fixed physical
distance from the host.

\begin{figure}
 \centering
 \hspace*{-0.12in}
 \includegraphics[width=3.45in]{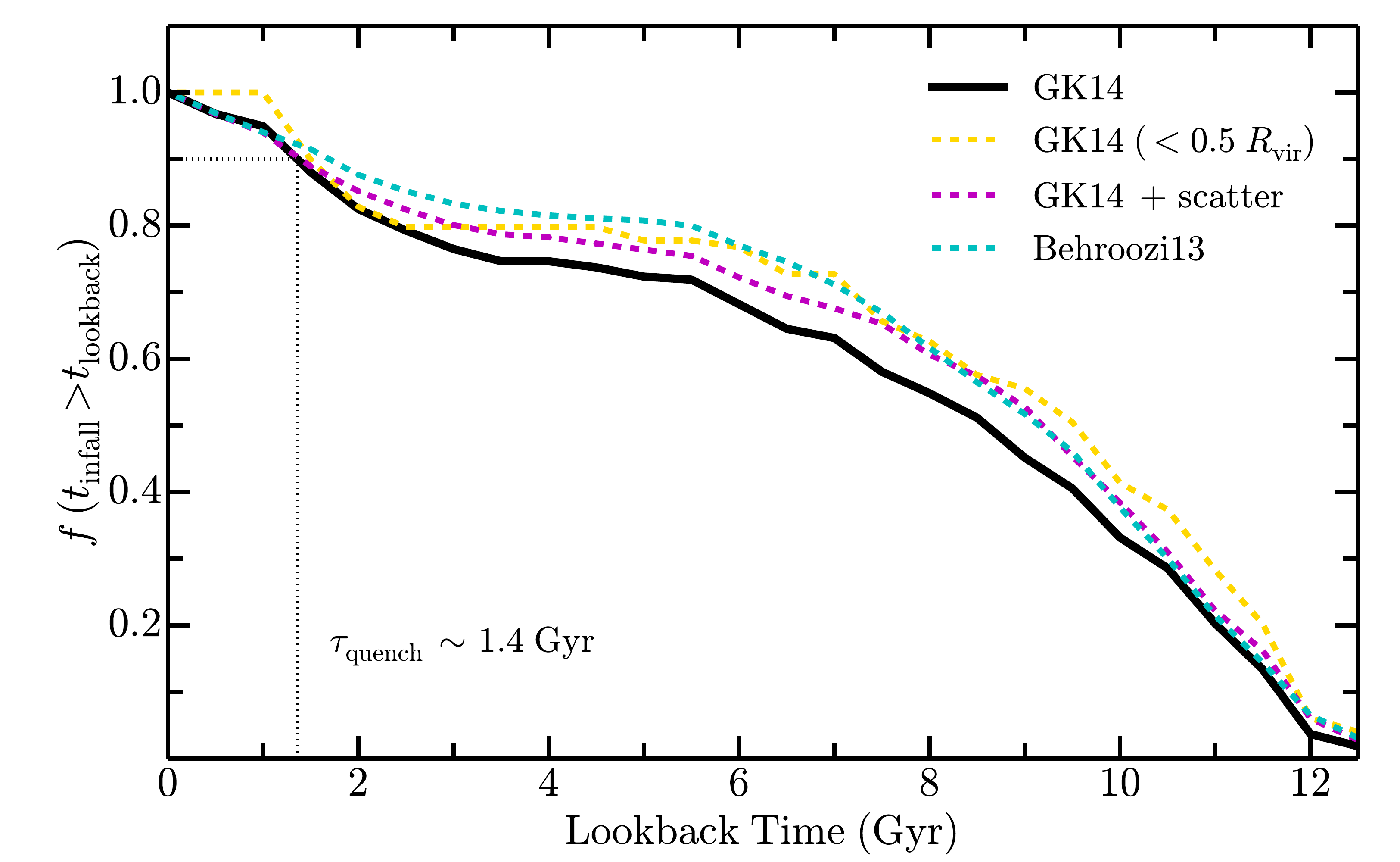}
 \caption{The cumulative distribution of infall times ($t_{\rm
     infall}$) as a function of lookback time for those subhalos in
   the ELVIS simulations identified as hosts of the Local Group
   satellites at $10^{6} < \mstar/\msun < 10^{8}$. The solid black
   line corresponds to our default selection criteria, where subhalos
   are selected according to the abundance matching relation of
   \citet{gk14}, restricting to those subhalos with $V_{\rm
     max}/V_{\rm peak} > 0.3$ and within the host virial radius at $z
   = 0$.  The dashed cyan and magenta lines show the corresponding
   distribution of infall times when applying alternative abundance
   matching prescriptions (see~\S\ref{sec:ELVIS}), while the dashed
   gold line traces the accretion history of only those subhalos
   within $0.5~R_{\rm vir}$ at $z = 0$ following the \citet{gk14}
   abundance matching relation. A quenched fraction of $90\%$ today,
   as observed for the Local Group at these masses, corresponds to a
   quenching timescale ($\tau_{\rm quench}$) of $\sim 1.4$~Gyr. }
 \label{fig:tinfall}
\end{figure}


\begin{figure*}
  \centering
  \hspace*{-0.42in}
  \includegraphics[width=7.5in]{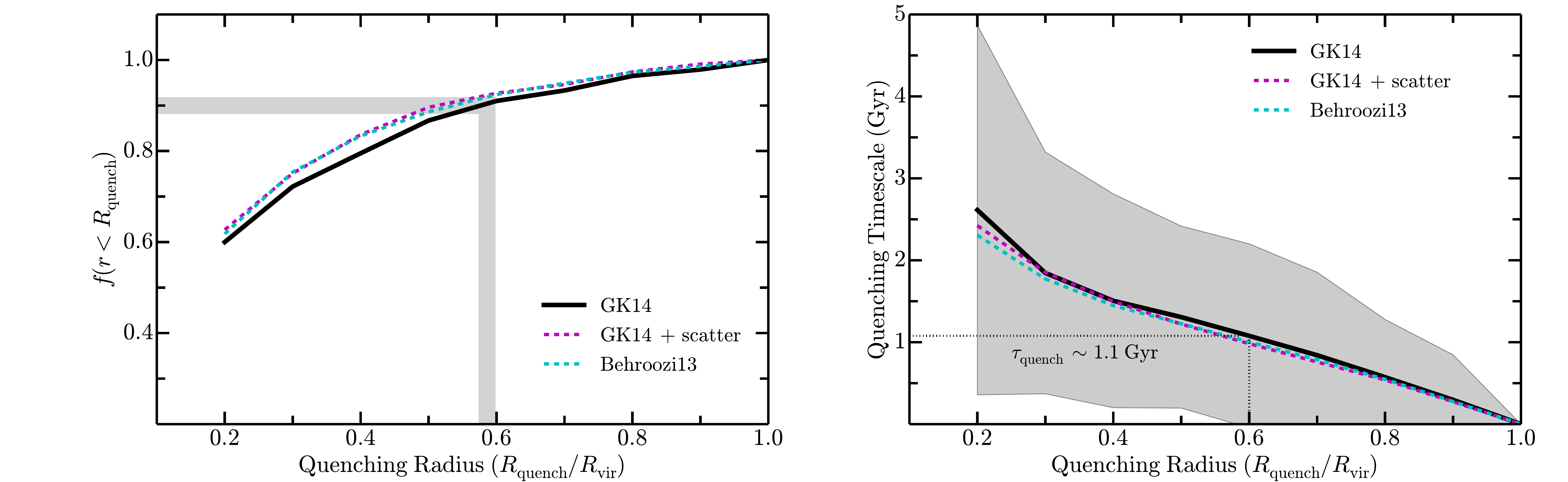}
  \caption{\emph{Left}: The dependence of the satellite quenched
    fraction on the assumed quenching radius, $R_{\rm quench}$, for our
    ram-pressure stripping model (see~\S\ref{subsec:rquench}). The
    solid black and dashed cyan and magenta lines correspond to subhalo
    populations selected according to different abundance matching
    prescriptions (see~\S\ref{sec:ELVIS}). In all cases, a quenching
    radius of $0.6~R_{\rm vir}$ is required to reproduce a quenched
    fraction of $90\%$ as observed in the Local Group. \emph{Right}:
    The dependence of the mean quenching timescale ($\tau_{\rm
      quench}$) on the assumed quenching radius ($R_{\rm quench}$) for
    the same set of abundance matching relations. For our fiducial
    subhalo sample, the grey shaded region illustrates the $1\sigma$
    scatter in the quenching timescale across the $48$ ELVIS hosts. As
    for the starvation model, the high observed quenched fraction in
    the Local Group implies a short quenching timescale ($\sim1.1$~Gyr)
    in our ram-pressure stripping model. The typical $R_{\rm vir}$ (at
    $z=0$) for ELVIS hosts is $\sim300$~kpc.}
  \label{fig:rquench}
\end{figure*}

\subsection{Starvation Model}
\label{subsec:tquench}

When quenching occurs via starvation, the satellite galaxy is deprived
of any additional supply of cold gas after infall, such that star
formation is halted following the consumption of its current gas
reservoir --- i.e.~quenching occurs within the gas depletion timescale
($\tau_{\rm depl} = {\rm M}_{\rm gas} / {\rm SFR}$). While
observations suggest that atomic and/or molecular depletion timescales
may vary with galaxy mass \citep{leroy08, bigiel11, boselli14,
  somerville15}, we adopt a fixed quenching timescale (following infall)
over the limited stellar mass range probed in our analysis
($10^{6}-10^{8}~\msun$).

In the models, we define the infall time as the lookback time at which
a halo \emph{first} crosses the virial radius, $R_{\rm vir}(z)$, of
the host. For the $\sim 23\%$ of subhalos that cross the host virial
radius more than once, we are likely overestimating the infall
time. However, $90\%$ of these subhalos are accreted for a final time
$\gtrsim6$~Gyr ago, thereby contributing little to no effect on the
final inferred quenching timescale. Figure~\ref{fig:tinfall} shows the
cumulative distribution of infall times for subhalos selected as hosts
of Local Group satellites in the simulations following various
abundance matching prescriptions. Universally, we find that subhalos
are accreted relatively uniformly over cosmic time, with the vast
majority infalling into their host halo $>2$~Gyr ago.

As shown in Figure~\ref{fig:tinfall}, to match the observed satellite
quenched fraction of $\sim90\%$ in the Local Group, the starvation
model -- independent of the adopted abundance matching prescription --
favors a quenching timescale ($\tau_{\rm quench}$) of roughly
$1.4^{+0.9}_{-0.7}$~Gyr, including the error associated with the
uncertainty in the observed Local Group quenched fraction (according
to binomial statistics). Thus, on average, satellites of the Milky Way
and M31 (at $\mstar \sim 10^{6}-10^{8}~\msun$) must quench within
$1.4$~Gyr of infalling onto their respective host halo. Within the
individual ELVIS simulations there is non-negligible scatter in the
derived quenching timescale, given the stochasticity of subhalo
accretion events; assuming a quenched fraction of $90\%$, the
$1\sigma$ scatter in $\tau_{\rm quench}$ across the $12$ Local
Group-like host pairs in ELVIS is approximately $0.7$~Gyr.

A quenched fraction of $90\%$ in the local group could be dependent on
completeness of the dwarf sample in this mass range, as discussed in
the introduction. This is particularly true around M31, where PAndAS
imaging only surveys the volume within a radius of $\sim150$~kpc. To
test the impact of this potential selection effect on our results, we
also compute the distribution of infall times for those subhalos that
reside within $0.5~R_{\rm vir}$ today. While simulated subhalos with
earlier infall times are slightly biased towards smaller host-centric
distances \citep[e.g.][]{diemand08, rocha12, wetzel15}, the
correlation between radial distance from the host and infall time
exhibits significant scatter, such that the quenching timescale
derived from subhalos within $0.5~R_{\rm vir}$ today is fully
consistent with that based on the full sample of subhalos within
$1~R_{\rm vir}$ (see~Fig.~\ref{fig:tinfall}).

In our starvation model, it is assumed that the quenching timescale
(i.e.~depletion timescale) is independent of cosmic time. While
studies at higher redshift are limited to significantly more massive
systems, current work suggests that molecular depletion timescales are
relatively weakly dependent on redshift \citep{geach11, magdis12a,
  saintogne13, bauermeister13}, favoring shorter timescales at higher
$z$. This conclusion is echoed in analysis of the satellite quenching
timescale across a broad range of halo masses at intermediate redshift
\citep{mcgee11, mcgee14, mok14, muzzin14, bahe15}. Altogether, a
modest variation in the quenching timescale with redshift would yield
no significant impact on our results, as our analysis is focused on
the cumulative quenched fraction today and not on the detailed
quenching times of individual satellites.


\subsection{Stripping Model}
\label{subsec:rquench}

To mimic quenching via ram-pressure stripping, we implement a model in
which satellites are quenched at a fixed radial distance ($R_{\rm
  quench}$) from their host. In a stripping scenario, this radial
quenching scale would be set by the volume over which the host's hot
halo reaches a density capable of stripping an infalling dwarf. To
account for growth of the hot halo over cosmic time, we define our
quenching radius, $R_{\rm quench}$, in terms of the host's virial
radius as a function of redshift (i.e.~$R_{\rm quench} \propto R_{\rm
  vir}(z)$) and require that a subhalo cross within the quenching
radius at $z < 3$, allowing the host's hot halo time to form
\citep{birnboim03, keres05, dekel06}. For subhalos in ELVIS that have
crossed within the quenching radius at $z < 3$, we classify the
corresponding satellite as quenched; all other systems remain
star-forming, independent of the time spent in the host halo
(i.e.~independent of infall time).

Figure~\ref{fig:rquench}a shows the dependence of the satellite
quenched fraction on the adopted quenching radius (from $0.2~R_{\rm
  vir}$ to $1~R_{\rm vir}$). We find that to reproduce the observed
quenched fraction in the Local Group ($90\%$) requires a quenching
radius of roughly $0.6~R_{\rm vir}$, again with little dependence on
the assumed abundance matching relation. Adopting this value for the
quenching radius ($R_{\rm quench} = 0.6~R_{\rm vir}$), we are able to
determine the time at which each subhalo in the simulation
(i.e.~satellite) quenched following infall --- and thus the quenching
timescale (measured relative to infall). In Figure~\ref{fig:rquench}b,
we show the mean quenching timescale as a function of the chosen
quenching radius ($R_{\rm quench}$) within the $48$ ELVIS
simulations. For large values of $R_{\rm quench}$, each satellite --
largely independent of orbit -- spends little time in the host halo
prior to being quenched, such that the effective quenching timescale
($\tau_{\rm quench}$) is short. Adopting a quenching radius of $R_{\rm
  quench} = 0.6~R_{\rm vir}$, we find an average quenching timescale
of $1.1$~Gyr, in very good agreement with that derived from our
starvation model (see \S\ref{subsec:tquench}). Finally, a
complementary analysis by \citet{slater14}, in which quenching
commences at pericenter passage, finds very similar results
\citep[$\tau_{\rm quench} \sim 1-2$~Gyr, see also][]{weisz15, wetzel15b}.


\section{Discussion}
\label{sec:discussion}

Given that nearly all field dwarfs are star-forming at stellar masses
of $\lesssim 10^{9}~\msun$ \citep{grebel97, mateo98, geha12},
environment-dependent processes are the dominant driver of quenching
at low masses and must act on a timescale so as to avoid over- or
under-producing the number of star-forming satellite galaxies in the
Local Group. Independent of the particular physical mechanism(s) at
play, the high quenched fraction ($\sim90\%$) observed for low-mass
Local Group satellites thereby dictates a remarkably short quenching
timescale, measured relative to infall. Through a detailed comparison
of the observed Milky Way and M31 satellites to the ELVIS suite of
simulations, we show that this quenching timescale ($\tau_{\rm
  quench}$) is $\sim 2$~Gyr at stellar masses of
$10^{6}-10^{8}~\msun$ in the Local Group.

A very similar analysis by \citet{wheeler14}, in which the properties
of nearby dwarfs in the SDSS are compared to the Millennium-II
simulation \citep{bk09}, finds that the quenching timescale must be
much longer ($>7.5$~Gyr) at slightly higher stellar masses
($\sim10^{9}~\msun$). This longer quenching timescale is dictated by
the relatively low quenched fractions observed in nearby groups
\citep{geha12}. While \citet{wheeler14} preferentially study
higher-mass host halos than that of the Milky Way and M31, analysis of
the Local Group is in broad agreement. At $\mstar \sim
10^{8}-10^{10}~\msun$, the satellite quenched fraction in the Local
Group is only $\sim40\%$ \citep{wheeler14, phillips15, slater14},
which requires a quenching timescale of $\sim7.8$~Gyr when comparing
the infall times of ELVIS subhalos selected to have $\mstar \sim
10^{8.5}-10^{9.5}~\msun$ according to the abundance matching relation
of \cite{gk14}.

\begin{figure*}
\centering
\hspace*{-0.15in}
\includegraphics[width=5.0in]{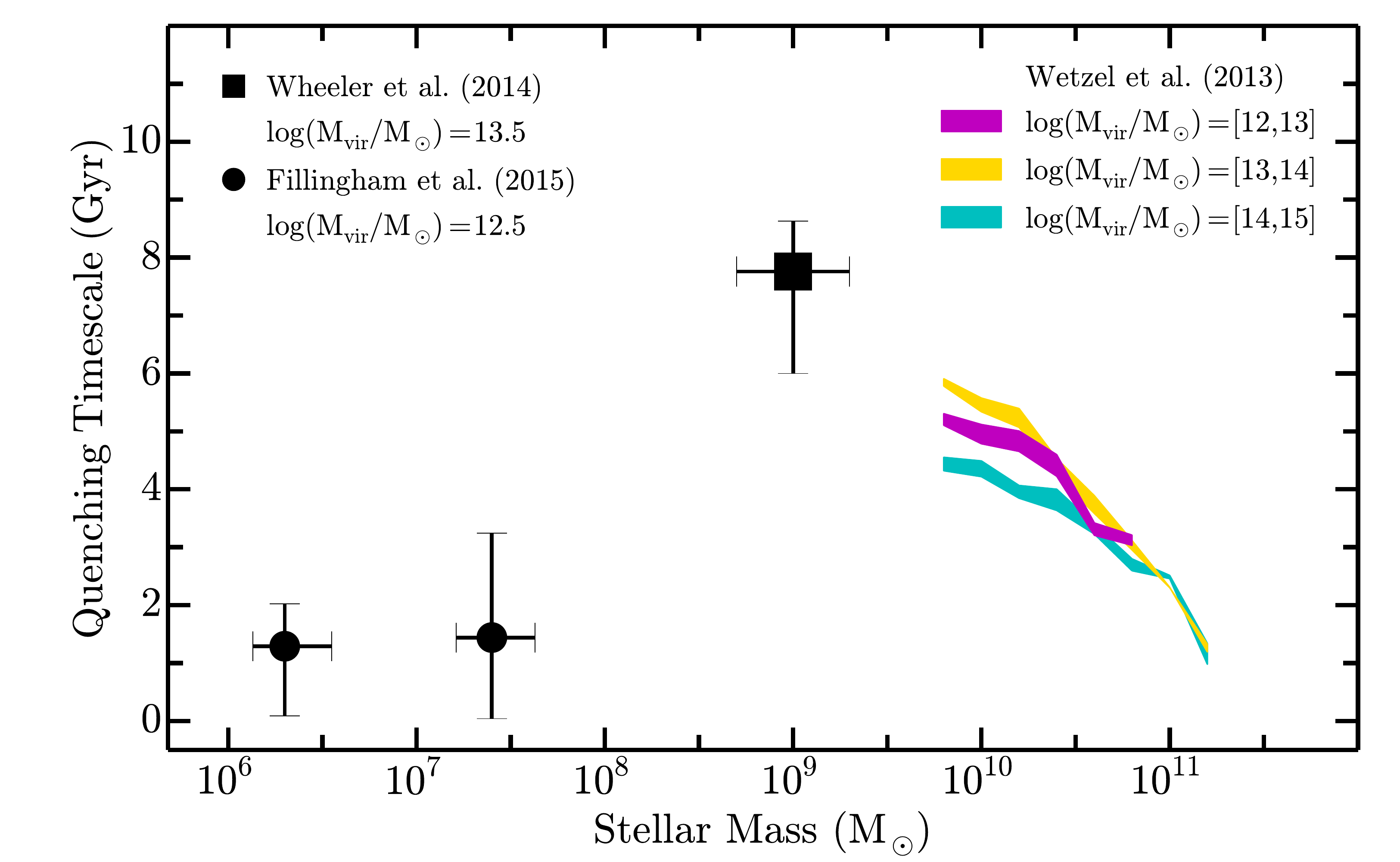}
\caption{The dependence of the satellite quenching timescale
  ($\tau_{\rm quench}$) on satellite stellar mass at $\mstar \sim
  10^{6}-10^{11}~\msun$. The inset legend lists the typical host halo
  mass for each data set. The magenta, gold, and cyan colored bands
  show the constraints for satellites in different mass host halos
  derived from analysis of galaxy groups and clusters in the SDSS by
  \citet{wetzel13}. The black square denotes the typical quenching
  timescale for slightly lower-mass satellites ($\sim10^{9}~\msun$)
  from \citet{wheeler14}, with the horizontal error bars denoting the
  $25-75\%$ range in stellar mass probed by that work and the vertical
  error bars giving the variation in $\tau_{\rm quench}$ corresponding
  to satellite quenched fractions of $25-55\%$ (as derived from
  analysis of subhalo populations in ELVIS, see
  \S\ref{sec:discussion}). Our estimate for the low-mass satellite
  quenching timescale in the Local Group is given by the black circles
  ($\tau_{\rm quench} \sim 1.4$~Gyr, see \S\ref{subsec:tquench}),
  where our sample of subhalos is divided into two stellar mass bins
  of $10^{6}-10^{7}~\msun$ and $10^{7}-10^{8}~\msun$. Each point gives
  the quenching timescale for subhalos in that mass bin assuming a
  satellite quenched fraction of $90\%$, while the vertical error bars
  illustrate the variation in $\tau_{\rm quench}$ corresponding to
  satellite quenched fractions of $80-100\%$. Our results are largely
  independent of mass, with increased scatter at higher masses due to
  the smaller number (and thus increased stochasticity in the infall
  times) of massive subhalos. The horizontal error bars show the
  quartiles of the subhalo mass distribution within each bin. At high
  masses, the quenching timescale increases with decreasing stellar
  mass. Below ${\rm M}_{\star}\sim10^{8}~\msun$, we find that
  satellite quenching becomes dramatically more efficient, suggesting
  a likely change in the physical mechanisms at play.}
\label{fig:tquench}
\end{figure*}

In an effort to study quenching for the most massive satellite
systems, \citet{wetzel13} and \citet{delucia12} compare group and
cluster populations from the SDSS \citep{yang07} to $N$-body and
semi-analytic models, respectively. While analysis at these masses is
complicated by the fact that some isolated or field systems are
quenched independent of any environment-related effects, these
independent analyses conclude that the quenching timescale at
$\sim10^{10}-10^{11}~\msun$ increases with decreasing satellite
stellar mass, reaching as long as $\sim 5-6$~Gyr \citep[see
also][]{hirschmann14}. Interestingly, for these most massive
satellites, no evidence is found for variation in the quenching
timescale with host halo mass \citep[][but see also
\citealt{phillips15}]{wetzel13}.

\begin{figure*}
 \centering
 \hspace*{-0.48in}
   \includegraphics[width=7.6in]{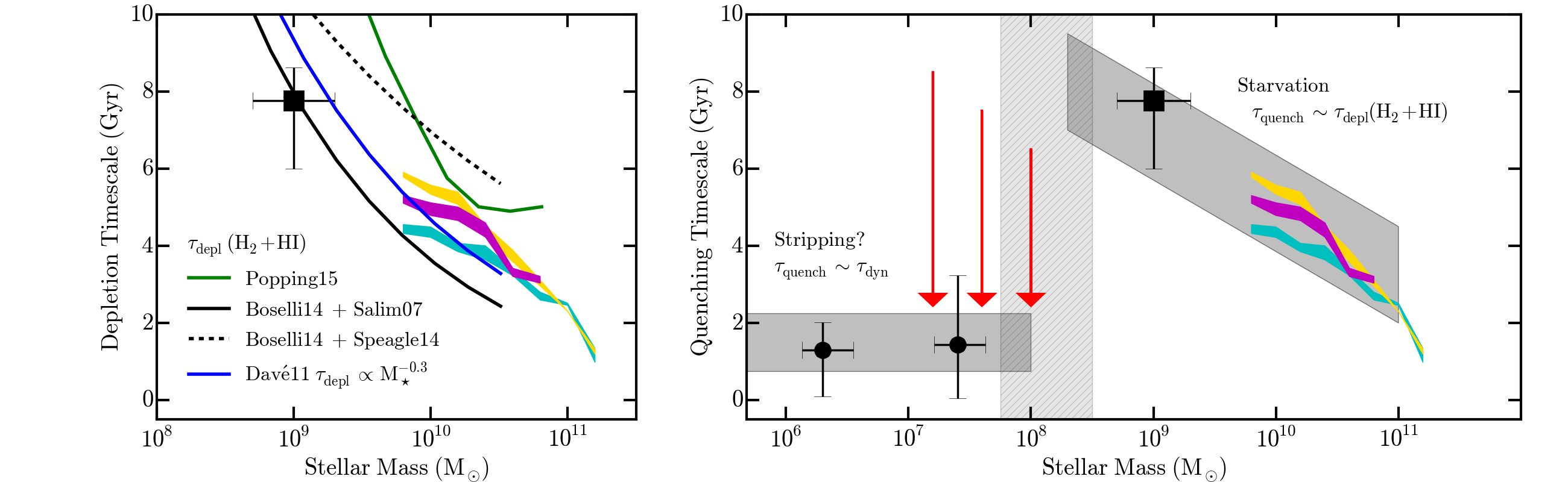}
   \caption{\emph{Left}: The dependence of the satellite quenching
     timescale ($\tau_{\rm quench}$) and gas depletion timescale
     ($\tau_{\rm depl}$) on satellite stellar mass at $\mstar \gtrsim
     10^{8}~\msun$. The magenta, gold, and cyan colored bands show the
     constraints for satellites in different mass host halos derived
     from analysis of galaxy groups and clusters in the SDSS by
     \citet{wetzel13}. The black square gives the typical quenching
     timescale for lower-mass satellites ($\sim
     10^{8.5}-10^{9.5}~\msun$) from \citet{wheeler14}. The solid and
     dashed black lines show the mass dependence of the gas depletion
     timescale (${\rm H}_{2}$ + H{\scriptsize I}) derived from
     combining the observed gas fractions of \citet{boselli14} with
     the star formation rate-stellar mass relation of \citet{salim07}
     and \citet{speagle14}, respectively. The solid blue and solid
     green lines give the corresponding predictions from the
     hydrodynamical and semi-empirical models of \citet{dave11a} and
     \citet{popping15}, respectively. The inferred satellite quenching
     timescales at $\gtrsim10^{8}~\msun$ show broad agreement with the
     observed gas depletion timescales for field systems, suggesting
     that starvation is the main driver of satellite quenching at
     these masses. \emph{Right}: The dependence of the satellite
     quenching timescale on satellite stellar mass at $10^{6} <
     \mstar/\msun < 10^{11}$, including the estimates of $\tau_{\rm
       quench}$ from Fig.~\ref{fig:tquench}. The light grey shaded
     regions highlight the expected dominant quenching mechanism and
     projected mass dependence of $\tau_{\rm quench}$, while the
     vertical hashed grey region demarcates the critical mass scale
     below which a more efficient physical process (e.g.~ram-pressure
     stripping) drives relatively rapid quenching in the Local
     Group. At high masses ($\gtrsim10^{8}~\msun$), satellite
     quenching is consistent with being driven by starvation; below
     $\sim10^{8}~\msun$, however, we posit that ram-pressure stripping
     becomes increasingly effective, such that depletion (and thus
     quenching) timescales are significantly shorter.}

 \label{fig:cartoon}
\end{figure*}

In Figure~\ref{fig:tquench}, we combine our estimate of the quenching
timescale for low-mass satellites in the Local Group ($\mstar~\sim
10^{7}~\msun$) with the complementary constraints from
\citet{wheeler14} and \citet{wetzel13} at higher stellar masses. Due
to differences in the adopted definiton of infall time (see discussion
in \citealt{wheeler14}), we assume a quenching timescale of $8$~Gyr at
$10^{9}~\msun$. This value for $\tau_{\rm quench}$ is fully consistent
with the results from \citet[][$\tau_{\rm
  quench}\sim7.5-9$~Gyr]{wheeler14}; as previously highlighted,
analysis of higher-mass subhalos in the ELVIS simulations --- systems
corresponding to $10^{8.5}-10^{9.5}~\msun$ following the \citet{gk14}
abundance matching relation --- yields a very similar result
($\tau_{\rm quench}\sim7.8$~Gyr; see also
\citealt{delucia12}). Altogether, we find that the quenching timescale
increases with decreasing mass at $\gtrsim10^{8}~\msun$, such that
environmental processes are relatively inefficient in suppressing star
formation at $\sim10^{9}~\msun$. At lower stellar masses
($\lesssim10^{8}~\msun$), however, the quenching timescale decreases
dramatically, indicating a stark increase in the satellite quenching
efficiency for low-mass systems. Given that all known satellites of
the Milky Way and M31 are quenched below stellar masses of $\sim{\rm
  a~few}~\times~10^{5}~\msun$, the quenching timescale is expected to
remain short to even lower masses.  For the lowest-mass systems
(i.e.~ultra-faint dwarfs) environmental quenching is likely
irrelevant, as the smallest dark matter halos may have their star
formation suppressed significantly by non-environmental mechanisms
like cosmic reionization \citep[e.g.][]{rees86, efstathiou92,
  wyithe06, onorbe15}.

For an infalling satellite of any mass, assumed to be isolated from
subsequent gas accretion, an upper limit on the longevity of any star
formation activity is set by the gas supply within the satellite upon
infall --- including the satellite's hot halo, from which gas may
cool. Processes such as ram-pressure stripping, tidal effects, and
harassment would then work to diminish this gas supply, so as to
accelerate the quenching process. The long quenching timescales
observed for higher-mass satellites ($\sim10^{8}-10^{10}~\msun$) thus
favor starvation as the dominant quenching mechanism. Ignoring future
accretion, recycling, or gas cooling, the duration of star formation
should follow the gas depletion timescale ($\tau_{\rm depl} = {\rm
  M}_{\rm gas} / {\rm SFR}$, where ${\rm M}_{\rm gas}$ is the mass of
a satellite's cold gas reservoir upon infall). While studies of nearby
star-forming galaxies suggest that the surface density of ongoing star
formation is most closely tied to the dense ${\rm H}_{2}$ gas
\citep{bigiel08, bigiel11, leroy08}, observed molecular depletion
timescales are much shorter than the inferred quenching
timescales. For relatively massive galaxies ($\mstar \gtrsim
10^{10}~\msun$), spanning a broad range of redshift, current
observations point towards molecular depletion timescales of roughly
$1-2$~Gyr \citep{bigiel11, leroy13, saintonge11, tacconi10, tacconi13,
  genzel10, genzel15}. Moreover, recent work targeting larger samples
of lower-mass galaxies conclude that the molecular depletion timescale
decreases with decreasing stellar mass, such that infalling satellites
at $10^{9}~\msun$ should, on average, exhaust their molecular
reservoirs in less than $1$~Gyr \citep{saintonge11, boselli14}.

These relatively short molecular depletion timescales are in direct
conflict with the derived quenching timescales for satellites at
$\sim10^{9}-10^{10}~\msun$, which indicate that star formation
typically continues for $>4$~Gyr after infall. Given that strong
variation in star-formation efficiency for ${\rm H}_{2}$ is disfavored
by both observations of the Kennicutt-Schmidt relation
\citep{bigiel08, leroy08, krumholz05} and the bimodality of rest-frame
galaxy colors \citep[or specific star formation rates,][]{balogh09,
  weinmann10},\footnote{While a reduction in the star formation
  efficiency for satellites would yield longer depletion (and thus
  quenching) timescales, it would also lead to overproducing galaxies
  in the ``green valley''.} the extended star formation in satellite
galaxies seemingly must be driven by a fuel supply beyond the
molecular reservoir. While gas cooling from a satellite's hot halo
could replenish the fuel supply for star formation, there is little
observational evidence for significant hot gas coronae around
star-forming galaxies, even at the high-mass regime and especially
when restricting to gas with a sufficiently short cooling time
\citep[][]{sun07, rasmussen09, crain10, miller15}.

On the other hand, if atomic gas is included as a potential fuel
source for star formation, there may be better agreement between the
expected gas depletion and satellite quenching timescales. As
previously stated, star formation is observed to follow molecular gas
density more closely than atomic gas density, although the weaker
correlation with H{\scriptsize I} surface density could be largely due
to atomic gas spanning a broader range of densities, thus on average
tracing dense star-forming clouds more poorly \citep{glover12a,
  glover12b, clark12}. As shown in Figure~\ref{fig:cartoon}a, when
including the atomic gas component, the inferred quenching timescales
at high stellar masses ($\mstar \gtrsim 10^{8}~\msun$) are in broad
agreement with the measured and predicted (atomic plus molecular) gas
depletion timescales and thus consistent with a picture where
satellite quenching is driven by starvation at high masses.

In the low-mass regime ($\lesssim10^{8}~\msun$), the atomic component
becomes increasingly important, with observations of nearby
star-forming systems finding that the typical H{\scriptsize I} gas
fraction increases steadily towards lower mass \citep{skillman03,
  geha06, leroy08, schiminovich10, catinella10, huang12}. More
importantly, this relative growth in atomic gas outpaces the
corresponding increase in the average specific star formation rate
\citep[e.g.][]{noeske07, salim07, speagle14}, such that the gas
depletion timescale is generally long ($\gtrsim 10$~Gyr) for low-mass
star-forming galaxies --- a result that is fully consistent with the
dearth of quiescent field galaxies at $<10^{9}~\msun$ \citep{geha12,
  onorbe15}. In comparison to the typical quenching timescales at low
masses, however, the overabundance of atomic gas in these systems
suggests that starvation alone cannot drive satellite quenching at
$\lesssim10^{8}~\msun$. Instead, a secondary process, such as
ram-pressure stripping, is needed to remove cool gas from infalling
systems, thereby decreasing the gas depletion timescale significantly.

As observations of systems such as Leo I suggest \citep{sohn13}, we
would expect stripping to occur as a satellite infalls from $R_{\rm
  vir}$ towards pericenter, such that quenching at low masses would
proceed according to the dynamical time ($\tau_{\rm dyn}$), which is
roughly $1-2$~Gyr for a Milky Way-like halo today
\citep[e.g.][]{stewart09}.
While a detailed analysis of the efficacy of ram-pressure stripping
(or other quenching mechanisms such as tidal forces) is beyond the
scope of this work (see Fillingham et al.~in prep),
Figure~\ref{fig:cartoon}b presents a qualitative depiction of the
potential physical processes at play, where satellite quenching is
primarily driven by starvation at higher masses ($\gtrsim
10^{8}~\msun$ for the Local Group) while ram-pressure stripping
becomes increasingly effective below a critical mass scale of $\sim
10^{8}~\msun$, yielding shorter quenching timescales and high observed
quenched fractions. In this qualitative picture, we expect that
ram-pressure stripping (or some other efficient quenching mechanism)
plays a critical role in suppressing star formation down to the mass
scales at which reionization inhibits gas accretion and therefore star
formation \citep[perhaps, $\mstar \sim 10^{4}~\msun$,][]{brown14,
  onorbe15, wheeler15}. Applying this qualitative picture to
higher-mass host halos (e.g.~rich groups and clusters), the critical
scale at which ram-pressure stripping becomes effective should move to
higher satellite masses due to the increased density of the host's hot
halo in concert with higher infall velocities for the satellites.

\section{Summary}
\label{sec:summary}

Comparing observations of the Local Group satellite population to
corresponding high-resolution $N$-body simulations from the ELVIS
suite, we investigate the typical timescale (following infall) upon
which star formation is suppressed in low-mass ($<10^{8}~\msun$)
satellite galaxies. When combining our work with complementary
analyses of higher-mass satellite populations, we present a
comprehensive picture of satellite quenching spanning roughly five
orders of magnitude in stellar mass. Our principal results are as
follows:

\begin{itemize}[leftmargin=0.25cm]

\item To reproduce the high fraction of quenched low-mass satellites
  in the Local Group, the typical quenching timescale at $\mstar \sim
  10^{6}-10^{8}~\msun$ must be relatively short ($\tau_{\rm
    quench} \sim 2$~Gyr).  \\

\item The longer quenching timescales inferred for higher-mass
  satellites are roughly consistent with the observed cold gas (${\rm
    H}_{2}$ plus H{\scriptsize I}) depletion timescales in
  corresponding field systems; this suggests that satellite quenching
  is largely driven by starvation at $\mstar \gtrsim 10^{8}~\msun$.  \\

\item At low masses ($\mstar \lesssim 10^{8}~\msun$), the much shorter
  satellite quenching timescales are potentially set by ram-pressure
  stripping or some other process that removes the cold gas reservoirs
  in infalling systems, such that nearly all low-mass satellites in
  the Local Group (or more massive host halos) are quenched.

\item If ram-pressure stripping is responsible for increasing the
  satellite quenching efficiency at low masses ($\mstar \lesssim
  10^{8}~\msun$) within the Local Group, we expect that this critical
  quenching scale will shift to higher satellite masses in higher-mass
  host halos (e.g.~rich groups and clusters) due to the increased
  density of the host's hot halo in concert with higher infall
  velocities for the satellites.

\end{itemize}

We have suggested here that the transition to rapid quenching in the
smallest satellites ($\mstar \lesssim 10^{8}~\msun$) is potentially
explained as a ram-pressure scale (see Figure~\ref{fig:cartoon} and
Fillingham et al.~in prep). Above this scale, satellite quenching
occurs on a gas depletion time, as accreted galaxies are starved of
fresh fuel for star formation.  Below this scale, a more rapid,
drag-induced stripping becomes dominant, such that the gravitational
restoring force of lower-mass dark matter halos can no longer retain
cool gas against the fluid pressure they experience while in orbit
around a larger host.

The idea that Local Group dwarfs are quenched via ram pressure has
been around for some time \citep[e.g.][]{einasto74, lin83, blitz00}
but the evidence for very rapid quenching provided here adds further
argument in its favor, and identifies a characteristic mass scale
where it may become dominant. Underlying this hypothesis is the notion
that the Milky Way and M31 both harbor extended ($\sim 150$ kpc)
reservoirs of hot baryons with sufficient density to strip small
galaxies of their gas. There is evidence that M31 does indeed host an
extended circumgalactic medium \citep{lehner15}, and X-ray studies are
consistent with this possibility around the Milky Way
\citep{fang13,miller15,fang15}. If extended low-density coronae of
this kind are common around $\sim L^{*}$ galaxies, then they may prove
important for understanding the baryon cycle and the global census of
baryons in the Universe.

Of course, one shortcoming of our analysis is that we have restricted
ourselves to satellites of M31 and the Milky Way. Given completeness
issues, the Local Group is a reasonable starting point, but it remains
possible that satellites in our vicinity are unusually quenched,
perhaps owing to an uncommonly dense distribution of hot baryons in
our vicinity, or atypically early infall times for the Local Group
satellite population. Ongoing efforts to discover and characterize
dwarf satellite systems around other massive hosts will be important
for solidifying the existence of a sharp transition scale at $\mstar
\lesssim 10^{8}~\msun$, where satellite quenching appears to become
extremely efficient in the Local Group. If future studies reveal that
the Local Group is indeed typical, then pinpointing the origin of this
quenching scale and testing the ram-pressure hypothesis will be of
utmost importance.

\section*{acknowledgements}

We thank Nicolas Martin and Andrew Wetzel for helpful discussions
regarding this work. We also thank the anonymous referee for useful
comments which helped clarify our work. CW and JSB were supported by
NSF grants AST-1009973 and AST-1009999. Support for this work was
provided by NASA through a Hubble Space Telescope theory grant
(program AR-12836) from the Space Telescope Science Institute (STScI),
which is operated by the Association of Universities for Research in
Astronomy (AURA), Inc., under NASA contract NAS5-26555. This research
made use of {\texttt{Astropy}}, a community-developed core Python
package for Astronomy \citep{astropy13}. Additionally, the Python
packages {\texttt{NumPy}} \citep{numpy}, {\texttt{iPython}}
\citep{ipython}, {\texttt{SciPy}} \citep{scipy}, and
{\texttt{matplotlib}} \citep{matplotlib} were utilized for the
majority of our data analysis and presentation.


\begin{thebibliography}{157}
\expandafter\ifx\csname natexlab\endcsname\relax\def\natexlab#1{#1}\fi

\bibitem[{{Astropy Collaboration} {et~al}\mbox{.}(2013){Astropy Collaboration},
  {Robitaille}, {Tollerud}, {Greenfield}, {Droettboom}, {Bray}, {Aldcroft},
  {Davis}, {Ginsburg}, {Price-Whelan}, {Kerzendorf}, {Conley}, {Crighton},
  {Barbary}, {Muna}, {Ferguson}, {Grollier}, {Parikh}, {Nair}, {Unther},
  {Deil}, {Woillez}, {Conseil}, {Kramer}, {Turner}, {Singer}, {Fox}, {Weaver},
  {Zabalza}, {Edwards}, {Azalee Bostroem}, {Burke}, {Casey}, {Crawford},
  {Dencheva}, {Ely}, {Jenness}, {Labrie}, {Lim}, {Pierfederici}, {Pontzen},
  {Ptak}, {Refsdal}, {Servillat}, \& {Streicher}}]{astropy13}
{Astropy Collaboration} {et~al.}, 2013, \aap, 558, A33

\bibitem[{{Bah{\'e}} \& {McCarthy}(2015)}]{bahe15}
{Bah{\'e}} Y.~M., {McCarthy} I.~G., 2015, \mnras, 447, 969

\bibitem[{{Baldry} {et~al}\mbox{.}(2006){Baldry}, {Balogh}, {Bower},
  {Glazebrook}, {Nichol}, {Bamford}, \& {Budavari}}]{baldry06}
{Baldry} I.~K., {Balogh} M.~L., {Bower} R.~G., {Glazebrook} K., {Nichol} R.~C.,
  {Bamford} S.~P., {Budavari} T., 2006, \mnras, 373, 469

\bibitem[{{Baldry} {et~al}\mbox{.}(2004){Baldry}, {Glazebrook}, {Brinkmann},
  {Ivezi{\'c}}, {Lupton}, {Nichol}, \& {Szalay}}]{baldry04}
{Baldry} I.~K., {Glazebrook} K., {Brinkmann} J., {Ivezi{\'c}} {\v Z}., {Lupton}
  R.~H., {Nichol} R.~C., {Szalay} A.~S., 2004, \apj, 600, 681

\bibitem[{{Balogh} {et~al}\mbox{.}(2004){Balogh}, {Baldry}, {Nichol}, {Miller},
  {Bower}, \& {Glazebrook}}]{balogh04}
{Balogh} M.~L., {Baldry} I.~K., {Nichol} R., {Miller} C., {Bower} R.,
  {Glazebrook} K., 2004, \apjl, 615, L101

\bibitem[{{Balogh} {et~al}\mbox{.}(2009){Balogh}, {McGee}, {Wilman}, {Bower},
  {Hau}, {Morris}, {Mulchaey}, {Oemler}, {Parker}, \& {Gwyn}}]{balogh09}
{Balogh} M.~L. {et~al.}, 2009, \mnras, 398, 754

\bibitem[{{Bauermeister} {et~al}\mbox{.}(2013){Bauermeister}, {Blitz},
  {Bolatto}, {Bureau}, {Leroy}, {Ostriker}, {Teuben}, {Wong}, \&
  {Wright}}]{bauermeister13}
{Bauermeister} A. {et~al.}, 2013, \apj, 768, 132

\bibitem[{{Bechtol} {et~al}\mbox{.}(2015){Bechtol}, {Drlica-Wagner},
  {Balbinot}, {Pieres}, {Simon}, {Yanny}, {Santiago}, {Wechsler}, {Frieman},
  {Walker}, {Williams}, {Rozo}, {Rykoff}, {Queiroz}, {Luque},
  {Benoit-L{\'e}vy}, {Tucker}, {Sevilla}, {Gruendl}, {da Costa}, {Fausti Neto},
  {Maia}, {Abbott}, {Allam}, {Armstrong}, {Bauer}, {Bernstein}, {Bernstein},
  {Bertin}, {Brooks}, {Buckley-Geer}, {Burke}, {Carnero Rosell}, {Castander},
  {Covarrubias}, {D'Andrea}, {DePoy}, {Desai}, {Diehl}, {Eifler}, {Estrada},
  {Evrard}, {Fernandez}, {Finley}, {Flaugher}, {Gaztanaga}, {Gerdes},
  {Girardi}, {Gladders}, {Gruen}, {Gutierrez}, {Hao}, {Honscheid}, {Jain},
  {James}, {Kent}, {Kron}, {Kuehn}, {Kuropatkin}, {Lahav}, {Li}, {Lin},
  {Makler}, {March}, {Marshall}, {Martini}, {Merritt}, {Miller}, {Miquel},
  {Mohr}, {Neilsen}, {Nichol}, {Nord}, {Ogando}, {Peoples}, {Petravick},
  {Plazas}, {Romer}, {Roodman}, {Sako}, {Sanchez}, {Scarpine}, {Schubnell},
  {Smith}, {Soares-Santos}, {Sobreira}, {Suchyta}, {Swanson}, {Tarle},
  {Thaler}, {Thomas}, {Wester}, {Zuntz}, \& {The DES Collaboration}}]{DES15}
{Bechtol} K. {et~al.}, 2015, \apj, 807, 50

\bibitem[{{Behroozi} {et~al}\mbox{.}(2013){Behroozi}, {Marchesini}, {Wechsler},
  {Muzzin}, {Papovich}, \& {Stefanon}}]{behroozi13}
{Behroozi} P.~S., {Marchesini} D., {Wechsler} R.~H., {Muzzin} A., {Papovich}
  C., {Stefanon} M., 2013, \apjl, 777, L10

\bibitem[{{Behroozi} {et~al}\mbox{.}(2014){Behroozi}, {Wechsler}, {Lu}, {Hahn},
  {Busha}, {Klypin}, \& {Primack}}]{behroozi14}
{Behroozi} P.~S., {Wechsler} R.~H., {Lu} Y., {Hahn} O., {Busha} M.~T., {Klypin}
  A., {Primack} J.~R., 2014, \apj, 787, 156

\bibitem[{{Bekki}(2009)}]{bekki09}
{Bekki} K., 2009, \mnras, 399, 2221

\bibitem[{{Bell} {et~al}\mbox{.}(2004){Bell}, {Wolf}, {Meisenheimer}, {Rix},
  {Borch}, {Dye}, {Kleinheinrich}, {Wisotzki}, \& {McIntosh}}]{bell04}
{Bell} E.~F. {et~al.}, 2004, \apj, 608, 752

\bibitem[{{Berrier} {et~al}\mbox{.}(2006){Berrier}, {Bullock}, {Barton},
  {Guenther}, {Zentner}, \& {Wechsler}}]{berrier06}
{Berrier} J.~C., {Bullock} J.~S., {Barton} E.~J., {Guenther} H.~D., {Zentner}
  A.~R., {Wechsler} R.~H., 2006, \apj, 652, 56

\bibitem[{{Bigiel} {et~al}\mbox{.}(2008){Bigiel}, {Leroy}, {Walter}, {Brinks},
  {de Blok}, {Madore}, \& {Thornley}}]{bigiel08}
{Bigiel} F., {Leroy} A., {Walter} F., {Brinks} E., {de Blok} W.~J.~G., {Madore}
  B., {Thornley} M.~D., 2008, \aj, 136, 2846

\bibitem[{{Bigiel} {et~al}\mbox{.}(2011){Bigiel}, {Leroy}, {Walter}, {Brinks},
  {de Blok}, {Kramer}, {Rix}, {Schruba}, {Schuster}, {Usero}, \&
  {Wiesemeyer}}]{bigiel11}
{Bigiel} F. {et~al.}, 2011, \apjl, 730, L13

\bibitem[{{Birnboim} \& {Dekel}(2003)}]{birnboim03}
{Birnboim} Y., {Dekel} A., 2003, \mnras, 345, 349

\bibitem[{{Blanton} {et~al}\mbox{.}(2005){Blanton}, {Eisenstein}, {Hogg},
  {Schlegel}, \& {Brinkmann}}]{blanton05}
{Blanton} M.~R., {Eisenstein} D., {Hogg} D.~W., {Schlegel} D.~J., {Brinkmann}
  J., 2005, \apj, 629, 143

\bibitem[{{Blitz} \& {Robishaw}(2000)}]{blitz00}
{Blitz} L., {Robishaw} T., 2000, \apj, 541, 675

\bibitem[{{Boselli} {et~al}\mbox{.}(2014){Boselli}, {Cortese}, {Boquien},
  {Boissier}, {Catinella}, {Gavazzi}, {Lagos}, \& {Saintonge}}]{boselli14}
{Boselli} A., {Cortese} L., {Boquien} M., {Boissier} S., {Catinella} B.,
  {Gavazzi} G., {Lagos} C., {Saintonge} A., 2014, \aap, 564, A67

\bibitem[{{Boylan-Kolchin}, {Bullock} \& {Kaplinghat}(2011){Boylan-Kolchin},
  {Bullock}, \& {Kaplinghat}}]{bk11}
{Boylan-Kolchin} M., {Bullock} J.~S., {Kaplinghat} M., 2011, \mnras, 415, L40

\bibitem[{{Boylan-Kolchin}, {Bullock} \& {Kaplinghat}(2012){Boylan-Kolchin},
  {Bullock}, \& {Kaplinghat}}]{bk12}
{Boylan-Kolchin} M., {Bullock} J.~S., {Kaplinghat} M., 2012, \mnras, 422, 1203

\bibitem[{{Boylan-Kolchin} {et~al}\mbox{.}(2013){Boylan-Kolchin}, {Bullock},
  {Sohn}, {Besla}, \& {van der Marel}}]{bk13}
{Boylan-Kolchin} M., {Bullock} J.~S., {Sohn} S.~T., {Besla} G., {van der Marel}
  R.~P., 2013, \apj, 768, 140

\bibitem[{{Boylan-Kolchin} {et~al}\mbox{.}(2009){Boylan-Kolchin}, {Springel},
  {White}, {Jenkins}, \& {Lemson}}]{bk09}
{Boylan-Kolchin} M., {Springel} V., {White} S.~D.~M., {Jenkins} A., {Lemson}
  G., 2009, \mnras, 398, 1150

\bibitem[{{Brasseur} {et~al}\mbox{.}(2011){Brasseur}, {Martin}, {Macci{\`o}},
  {Rix}, \& {Kang}}]{brasseur11}
{Brasseur} C.~M., {Martin} N.~F., {Macci{\`o}} A.~V., {Rix} H.-W., {Kang} X.,
  2011, \apj, 743, 179

\bibitem[{{Brown} {et~al}\mbox{.}(2007){Brown}, {Dey}, {Jannuzi}, {Brand},
  {Benson}, {Brodwin}, {Croton}, \& {Eisenhardt}}]{brown07}
{Brown} M.~J.~I., {Dey} A., {Jannuzi} B.~T., {Brand} K., {Benson} A.~J.,
  {Brodwin} M., {Croton} D.~J., {Eisenhardt} P.~R., 2007, \apj, 654, 858

\bibitem[{{Brown} {et~al}\mbox{.}(2014){Brown}, {Tumlinson}, {Geha}, {Simon},
  {Vargas}, {VandenBerg}, {Kirby}, {Kalirai}, {Avila}, {Gennaro}, {Ferguson},
  {Mu{\~n}oz}, {Guhathakurta}, \& {Renzini}}]{brown14}
{Brown} T.~M. {et~al.}, 2014, \apj, 796, 91

\bibitem[{{Bundy} {et~al}\mbox{.}(2006){Bundy}, {Ellis}, {Conselice}, {Taylor},
  {Cooper}, {Willmer}, {Weiner}, {Coil}, {Noeske}, \& {Eisenhardt}}]{bundy06}
{Bundy} K. {et~al.}, 2006, \apj, 651, 120

\bibitem[{{Burton} \& {Lockman}(1999)}]{burton99}
{Burton} W.~B., {Lockman} F.~J., 1999, \aap, 349, 7

\bibitem[{{Butler} {et~al}\mbox{.}(2007){Butler}, {Mart{\'{\i}}nez-Delgado},
  {Rix}, {Pe{\~n}arrubia}, \& {de Jong}}]{butler07}
{Butler} D.~J., {Mart{\'{\i}}nez-Delgado} D., {Rix} H.-W., {Pe{\~n}arrubia} J.,
  {de Jong} J.~T.~A., 2007, \aj, 133, 2274

\bibitem[{{Catinella} {et~al}\mbox{.}(2010){Catinella}, {Schiminovich},
  {Kauffmann}, {Fabello}, {Wang}, {Hummels}, {Lemonias}, {Moran}, {Wu},
  {Giovanelli}, {Haynes}, {Heckman}, {Basu-Zych}, {Blanton}, {Brinchmann},
  {Budav{\'a}ri}, {Gon{\c c}alves}, {Johnson}, {Kennicutt}, {Madore}, {Martin},
  {Rich}, {Tacconi}, {Thilker}, {Wild}, \& {Wyder}}]{catinella10}
{Catinella} B. {et~al.}, 2010, \mnras, 403, 683

\bibitem[{{Cheung} {et~al}\mbox{.}(2012){Cheung}, {Faber}, {Koo}, {Dutton},
  {Simard}, {McGrath}, {Huang}, {Bell}, {Dekel}, {Fang}, {Salim}, {Barro},
  {Bundy}, {Coil}, {Cooper}, {Conselice}, {Davis}, {Dom{\'{\i}}nguez},
  {Kassin}, {Kocevski}, {Koekemoer}, {Lin}, {Lotz}, {Newman}, {Phillips},
  {Rosario}, {Weiner}, \& {Willmer}}]{cheung12}
{Cheung} E. {et~al.}, 2012, \apj, 760, 131

\bibitem[{{Clark} {et~al}\mbox{.}(2012){Clark}, {Glover}, {Klessen}, \&
  {Bonnell}}]{clark12}
{Clark} P.~C., {Glover} S.~C.~O., {Klessen} R.~S., {Bonnell} I.~A., 2012,
  \mnras, 424, 2599

\bibitem[{{Conn} {et~al}\mbox{.}(2012){Conn}, {Ibata}, {Lewis}, {Parker},
  {Zucker}, {Martin}, {McConnachie}, {Irwin}, {Tanvir}, {Fardal}, {Ferguson},
  {Chapman}, \& {Valls-Gabaud}}]{conn12}
{Conn} A.~R. {et~al.}, 2012, \apj, 758, 11

\bibitem[{{Conroy}, {Wechsler} \& {Kravtsov}(2006){Conroy}, {Wechsler}, \&
  {Kravtsov}}]{conroy06}
{Conroy} C., {Wechsler} R.~H., {Kravtsov} A.~V., 2006, \apj, 647, 201

\bibitem[{{Cooper} {et~al}\mbox{.}(2010{\natexlab{a}}){Cooper}, {Coil},
  {Gerke}, {Newman}, {Bundy}, {Conselice}, {Croton}, {Davis}, {Faber},
  {Guhathakurta}, {Koo}, {Lin}, {Weiner}, {Willmer}, \& {Yan}}]{cooper10b}
{Cooper} M.~C. {et~al.}, 2010{\natexlab{a}}, \mnras, 409, 337

\bibitem[{{Cooper} {et~al}\mbox{.}(2010{\natexlab{b}}){Cooper}, {Gallazzi},
  {Newman}, \& {Yan}}]{cooper10a}
{Cooper} M.~C., {Gallazzi} A., {Newman} J.~A., {Yan} R., 2010{\natexlab{b}},
  \mnras, 402, 1942

\bibitem[{{Cooper} {et~al}\mbox{.}(2007){Cooper}, {Newman}, {Coil}, {Croton},
  {Gerke}, {Yan}, {Davis}, {Faber}, {Guhathakurta}, {Koo}, {Weiner}, \&
  {Willmer}}]{cooper07}
{Cooper} M.~C. {et~al.}, 2007, \mnras, 376, 1445

\bibitem[{{Cooper} {et~al}\mbox{.}(2006){Cooper}, {Newman}, {Croton}, {Weiner},
  {Willmer}, {Gerke}, {Madgwick}, {Faber}, {Davis}, {Coil}, {Finkbeiner},
  {Guhathakurta}, \& {Koo}}]{cooper06}
{Cooper} M.~C. {et~al.}, 2006, \mnras, 370, 198

\bibitem[{{Crain} {et~al}\mbox{.}(2010){Crain}, {McCarthy}, {Frenk}, {Theuns},
  \& {Schaye}}]{crain10}
{Crain} R.~A., {McCarthy} I.~G., {Frenk} C.~S., {Theuns} T., {Schaye} J., 2010,
  \mnras, 407, 1403

\bibitem[{{Dav{\'e}}, {Oppenheimer} \& {Finlator}(2011){Dav{\'e}},
  {Oppenheimer}, \& {Finlator}}]{dave11a}
{Dav{\'e}} R., {Oppenheimer} B.~D., {Finlator} K., 2011, \mnras, 415, 11

\bibitem[{{De Lucia} {et~al}\mbox{.}(2012){De Lucia}, {Weinmann}, {Poggianti},
  {Arag{\'o}n-Salamanca}, \& {Zaritsky}}]{delucia12}
{De Lucia} G., {Weinmann} S., {Poggianti} B.~M., {Arag{\'o}n-Salamanca} A.,
  {Zaritsky} D., 2012, \mnras, 423, 1277

\bibitem[{{Dekel} \& {Birnboim}(2006)}]{dekel06}
{Dekel} A., {Birnboim} Y., 2006, \mnras, 368, 2

\bibitem[{{Diehl} {et~al}\mbox{.}(2014){Diehl}, {Abbott}, {Annis}, {Armstrong},
  {Baruah}, {Bermeo}, {Bernstein}, {Beynon}, {Bruderer}, {Buckley-Geer},
  {Campbell}, {Capozzi}, {Carter}, {Casas}, {Clerkin}, {Covarrubias}, {Cuhna},
  {D'Andrea}, {da Costa}, {Das}, {DePoy}, {Dietrich}, {Drlica-Wagner},
  {Elliott}, {Eifler}, {Estrada}, {Etherington}, {Flaugher}, {Frieman}, {Fausti
  Neto}, {Gelman}, {Gerdes}, {Gruen}, {Gruendl}, {Hao}, {Head}, {Helsby},
  {Hoffman}, {Honscheid}, {James}, {Johnson}, {Kacprzac}, {Katsaros},
  {Kennedy}, {Kent}, {Kessler}, {Kim}, {Krause}, {Kron}, {Kuhlmann}, {Kunder},
  {Li}, {Lin}, {Maccrann}, {March}, {Marshall}, {Neilsen}, {Nugent}, {Martini},
  {Melchior}, {Menanteau}, {Nichol}, {Nord}, {Ogando}, {Old}, {Papadopoulos},
  {Patton}, {Petravick}, {Plazas}, {Poulton}, {Pujol}, {Reil}, {Rigby},
  {Romer}, {Roodman}, {Rooney}, {Sanchez Alvaro}, {Serrano}, {Sheldon},
  {Smith}, {Smith}, {Soares-Santos}, {Soumagnac}, {Spinka}, {Suchyta},
  {Tucker}, {Walker}, {Wester}, {Wiesner}, {Wilcox}, {Williams}, {Yanny}, \&
  {Zhang}}]{DES14}
{Diehl} H.~T. {et~al.}, 2014, in Society of Photo-Optical Instrumentation
  Engineers (SPIE) Conference Series, Vol. 9149, Society of Photo-Optical
  Instrumentation Engineers (SPIE) Conference Series, p.~0

\bibitem[{{Diemand} \& {Kuhlen}(2008)}]{diemand08}
{Diemand} J., {Kuhlen} M., 2008, \apjl, 680, L25

\bibitem[{{Efstathiou}(1992)}]{efstathiou92}
{Efstathiou} G., 1992, \mnras, 256, 43P

\bibitem[{{Einasto} {et~al}\mbox{.}(1974){Einasto}, {Saar}, {Kaasik}, \&
  {Chernin}}]{einasto74}
{Einasto} J., {Saar} E., {Kaasik} A., {Chernin} A.~D., 1974, \nat, 252, 111

\bibitem[{{Faber} {et~al}\mbox{.}(2007){Faber}, {Willmer}, {Wolf}, {Koo},
  {Weiner}, {Newman}, {Im}, {Coil}, {Conroy}, {Cooper}, {Davis}, {Finkbeiner},
  {Gerke}, {Gebhardt}, {Groth}, {Guhathakurta}, {Harker}, {Kaiser}, {Kassin},
  {Kleinheinrich}, {Konidaris}, {Kron}, {Lin}, {Luppino}, {Madgwick},
  {Meisenheimer}, {Noeske}, {Phillips}, {Sarajedini}, {Schiavon}, {Simard},
  {Szalay}, {Vogt}, \& {Yan}}]{faber07}
{Faber} S.~M. {et~al.}, 2007, \apj, 665, 265

\bibitem[{{Fang}, {Bullock} \& {Boylan-Kolchin}(2013){Fang}, {Bullock}, \&
  {Boylan-Kolchin}}]{fang13}
{Fang} T., {Bullock} J., {Boylan-Kolchin} M., 2013, \apj, 762, 20

\bibitem[{{Fang} {et~al}\mbox{.}(2015){Fang}, {Buote}, {Bullock}, \&
  {Ma}}]{fang15}
{Fang} T., {Buote} D., {Bullock} J., {Ma} R., 2015, \apjs, 217, 21

\bibitem[{{Ferguson} {et~al}\mbox{.}(2002){Ferguson}, {Irwin}, {Ibata},
  {Lewis}, \& {Tanvir}}]{ferguson02}
{Ferguson} A.~M.~N., {Irwin} M.~J., {Ibata} R.~A., {Lewis} G.~F., {Tanvir}
  N.~R., 2002, \aj, 124, 1452

\bibitem[{{Garrison-Kimmel} {et~al}\mbox{.}(2014b){Garrison-Kimmel},
  {Boylan-Kolchin}, {Bullock}, \& {Kirby}}]{gk14b}
{Garrison-Kimmel} S., {Boylan-Kolchin} M., {Bullock} J.~S., {Kirby} E.~N.,
  2014b, \mnras, 444, 222

\bibitem[{{Garrison-Kimmel} {et~al}\mbox{.}(2014a){Garrison-Kimmel},
  {Boylan-Kolchin}, {Bullock}, \& {Lee}}]{gk14}
{Garrison-Kimmel} S., {Boylan-Kolchin} M., {Bullock} J.~S., {Lee} K., 2014a,
  \mnras, 438, 2578

\bibitem[{{Geach} {et~al}\mbox{.}(2011){Geach}, {Smail}, {Moran}, {MacArthur},
  {Lagos}, \& {Edge}}]{geach11}
{Geach} J.~E., {Smail} I., {Moran} S.~M., {MacArthur} L.~A., {Lagos} C.~d.~P.,
  {Edge} A.~C., 2011, \apjl, 730, L19

\bibitem[{{Geha} {et~al}\mbox{.}(2006){Geha}, {Blanton}, {Masjedi}, \&
  {West}}]{geha06}
{Geha} M., {Blanton} M.~R., {Masjedi} M., {West} A.~A., 2006, \apj, 653, 240

\bibitem[{{Geha} {et~al}\mbox{.}(2012){Geha}, {Blanton}, {Yan}, \&
  {Tinker}}]{geha12}
{Geha} M., {Blanton} M.~R., {Yan} R., {Tinker} J.~L., 2012, \apj, 757, 85

\bibitem[{{Genzel} {et~al}\mbox{.}(2010){Genzel}, {Tacconi}, {Gracia-Carpio},
  {Sternberg}, {Cooper}, {Shapiro}, {Bolatto}, {Bouch{\'e}}, {Bournaud},
  {Burkert}, {Combes}, {Comerford}, {Cox}, {Davis}, {Schreiber},
  {Garcia-Burillo}, {Lutz}, {Naab}, {Neri}, {Omont}, {Shapley}, \&
  {Weiner}}]{genzel10}
{Genzel} R. {et~al.}, 2010, \mnras, 407, 2091

\bibitem[{{Genzel} {et~al}\mbox{.}(2015){Genzel}, {Tacconi}, {Lutz},
  {Saintonge}, {Berta}, {Magnelli}, {Combes}, {Garc{\'{\i}}a-Burillo}, {Neri},
  {Bolatto}, {Contini}, {Lilly}, {Boissier}, {Boone}, {Bouch{\'e}}, {Bournaud},
  {Burkert}, {Carollo}, {Colina}, {Cooper}, {Cox}, {Feruglio}, {F{\"o}rster
  Schreiber}, {Freundlich}, {Gracia-Carpio}, {Juneau}, {Kovac}, {Lippa},
  {Naab}, {Salome}, {Renzini}, {Sternberg}, {Walter}, {Weiner}, {Weiss}, \&
  {Wuyts}}]{genzel15}
{Genzel} R. {et~al.}, 2015, \apj, 800, 20

\bibitem[{{Giovanelli} {et~al}\mbox{.}(2013){Giovanelli}, {Haynes}, {Adams},
  {Cannon}, {Rhode}, {Salzer}, {Skillman}, {Bernstein-Cooper}, \&
  {McQuinn}}]{giovanelli13}
{Giovanelli} R. {et~al.}, 2013, \aj, 146, 15

\bibitem[{{Glover} \& {Clark}(2012{\natexlab{a}})}]{glover12a}
{Glover} S.~C.~O., {Clark} P.~C., 2012{\natexlab{a}}, \mnras, 421, 9

\bibitem[{{Glover} \& {Clark}(2012{\natexlab{b}})}]{glover12b}
{Glover} S.~C.~O., {Clark} P.~C., 2012{\natexlab{b}}, \mnras, 426, 377

\bibitem[{{Grcevich} \& {Putman}(2009)}]{grcevich09}
{Grcevich} J., {Putman} M.~E., 2009, \apj, 696, 385

\bibitem[{{Grebel}(1997)}]{grebel97}
{Grebel} E.~K., 1997, in Reviews in Modern Astronomy, Vol.~10, Reviews in
  Modern Astronomy, {Schielicke} R.~E., ed., pp. 29--60

\bibitem[{{Gunn} \& {Gott}(1972)}]{gunn72}
{Gunn} J.~E., {Gott}, III J.~R., 1972, \apj, 176, 1

\bibitem[{{Hirschmann} {et~al}\mbox{.}(2014){Hirschmann}, {De Lucia}, {Wilman},
  {Weinmann}, {Iovino}, {Cucciati}, {Zibetti}, \& {Villalobos}}]{hirschmann14}
{Hirschmann} M., {De Lucia} G., {Wilman} D., {Weinmann} S., {Iovino} A.,
  {Cucciati} O., {Zibetti} S., {Villalobos} {\'A}., 2014, \mnras, 444, 2938

\bibitem[{{Huang} {et~al}\mbox{.}(2012{\natexlab{a}}){Huang}, {Haynes},
  {Giovanelli}, \& {Brinchmann}}]{huang12}
{Huang} S., {Haynes} M.~P., {Giovanelli} R., {Brinchmann} J.,
  2012{\natexlab{a}}, \apj, 756, 113

\bibitem[{{Huang} {et~al}\mbox{.}(2012{\natexlab{b}}){Huang}, {Haynes},
  {Giovanelli}, {Brinchmann}, {Stierwalt}, \& {Neff}}]{huang12a}
{Huang} S., {Haynes} M.~P., {Giovanelli} R., {Brinchmann} J., {Stierwalt} S.,
  {Neff} S.~G., 2012{\natexlab{b}}, \aj, 143, 133

\bibitem[{Hunter(2007)}]{matplotlib}
Hunter J.~D., 2007, Computing In Science \& Engineering, 9, 90

\bibitem[{{Ibata} {et~al}\mbox{.}(2007){Ibata}, {Martin}, {Irwin}, {Chapman},
  {Ferguson}, {Lewis}, \& {McConnachie}}]{ibata07}
{Ibata} R., {Martin} N.~F., {Irwin} M., {Chapman} S., {Ferguson} A.~M.~N.,
  {Lewis} G.~F., {McConnachie} A.~W., 2007, \apj, 671, 1591

\bibitem[{{Irwin}(1994)}]{irwin94}
{Irwin} M.~J., 1994, in European Southern Observatory Conference and Workshop
  Proceedings, Vol.~49, European Southern Observatory Conference and Workshop
  Proceedings, {Meylan} G., {Prugniel} P., eds., p.~27

\bibitem[{Jones {et~al}\mbox{.}(2001--)Jones, Oliphant, Peterson,
  {et~al.}}]{scipy}
Jones E., Oliphant T., Peterson P., {et~al.}, 2001--, {SciPy}: Open source
  scientific tools for {Python}. [Online; accessed 2015-08-25]

\bibitem[{{Kaiser} {et~al}\mbox{.}(2010){Kaiser}, {Burgett}, {Chambers},
  {Denneau}, {Heasley}, {Jedicke}, {Magnier}, {Morgan}, {Onaka}, \&
  {Tonry}}]{kaiser10}
{Kaiser} N. {et~al.}, 2010, in Society of Photo-Optical Instrumentation
  Engineers (SPIE) Conference Series, Vol. 7733, Society of Photo-Optical
  Instrumentation Engineers (SPIE) Conference Series

\bibitem[{{Kaisin} \& {Karachentsev}(2013)}]{kaisin13}
{Kaisin} S.~S., {Karachentsev} I.~D., 2013, Astrophysics, 56, 305

\bibitem[{{Kaisin}, {Karachentsev} \& {Ravindranath}(2012){Kaisin},
  {Karachentsev}, \& {Ravindranath}}]{kaisin12}
{Kaisin} S.~S., {Karachentsev} I.~D., {Ravindranath} S., 2012, \mnras, 425,
  2083

\bibitem[{{Karachentsev} \& {Kaisina}(2013)}]{karachentsev13}
{Karachentsev} I.~D., {Kaisina} E.~I., 2013, \aj, 146, 46

\bibitem[{{Kauffmann} {et~al}\mbox{.}(2004){Kauffmann}, {White}, {Heckman},
  {M{\'e}nard}, {Brinchmann}, {Charlot}, {Tremonti}, \&
  {Brinkmann}}]{kauffmann04}
{Kauffmann} G., {White} S.~D.~M., {Heckman} T.~M., {M{\'e}nard} B.,
  {Brinchmann} J., {Charlot} S., {Tremonti} C., {Brinkmann} J., 2004, \mnras,
  353, 713

\bibitem[{{Kawata} \& {Mulchaey}(2008)}]{kawata08}
{Kawata} D., {Mulchaey} J.~S., 2008, \apjl, 672, L103

\bibitem[{{Kere{\v s}} {et~al}\mbox{.}(2005){Kere{\v s}}, {Katz}, {Weinberg},
  \& {Dav{\'e}}}]{keres05}
{Kere{\v s}} D., {Katz} N., {Weinberg} D.~H., {Dav{\'e}} R., 2005, \mnras, 363,
  2

\bibitem[{{Kimm} {et~al}\mbox{.}(2009){Kimm}, {Somerville}, {Yi}, {van den
  Bosch}, {Salim}, {Fontanot}, {Monaco}, {Mo}, {Pasquali}, {Rich}, \&
  {Yang}}]{kimm09}
{Kimm} T. {et~al.}, 2009, \mnras, 394, 1131

\bibitem[{{Kirby} {et~al}\mbox{.}(2014){Kirby}, {Bullock}, {Boylan-Kolchin},
  {Kaplinghat}, \& {Cohen}}]{kirby14}
{Kirby} E.~N., {Bullock} J.~S., {Boylan-Kolchin} M., {Kaplinghat} M., {Cohen}
  J.~G., 2014, \mnras, 439, 1015

\bibitem[{{Komatsu} {et~al}\mbox{.}(2011){Komatsu}, {Smith}, {Dunkley},
  {Bennett}, {Gold}, {Hinshaw}, {Jarosik}, {Larson}, {Nolta}, {Page},
  {Spergel}, {Halpern}, {Hill}, {Kogut}, {Limon}, {Meyer}, {Odegard}, {Tucker},
  {Weiland}, {Wollack}, \& {Wright}}]{komatsu11}
{Komatsu} E. {et~al.}, 2011, \apjs, 192, 18

\bibitem[{{Koposov} {et~al}\mbox{.}(2008){Koposov}, {Belokurov}, {Evans},
  {Hewett}, {Irwin}, {Gilmore}, {Zucker}, {Rix}, {Fellhauer}, {Bell}, \&
  {Glushkova}}]{koposov08}
{Koposov} S. {et~al.}, 2008, \apj, 686, 279

\bibitem[{{Koposov} {et~al}\mbox{.}(2015){Koposov}, {Belokurov}, {Torrealba},
  \& {Evans}}]{koposov15}
{Koposov} S.~E., {Belokurov} V., {Torrealba} G., {Evans} N.~W., 2015, \apj,
  805, 130

\bibitem[{{Krumholz} \& {McKee}(2005)}]{krumholz05}
{Krumholz} M.~R., {McKee} C.~F., 2005, \apj, 630, 250

\bibitem[{{Laevens} {et~al}\mbox{.}(2014){Laevens}, {Martin}, {Sesar},
  {Bernard}, {Rix}, {Slater}, {Bell}, {Ferguson}, {Schlafly}, {Burgett},
  {Chambers}, {Denneau}, {Draper}, {Kaiser}, {Kudritzki}, {Magnier},
  {Metcalfe}, {Morgan}, {Price}, {Sweeney}, {Tonry}, {Wainscoat}, \&
  {Waters}}]{laevens14}
{Laevens} B.~P.~M. {et~al.}, 2014, \apjl, 786, L3

\bibitem[{{Larson}, {Tinsley} \& {Caldwell}(1980){Larson}, {Tinsley}, \&
  {Caldwell}}]{larson80}
{Larson} R.~B., {Tinsley} B.~M., {Caldwell} C.~N., 1980, \apj, 237, 692

\bibitem[{{Lehner}, {Howk} \& {Wakker}(2015){Lehner}, {Howk}, \&
  {Wakker}}]{lehner15}
{Lehner} N., {Howk} J.~C., {Wakker} B.~P., 2015, \apj, 804, 79

\bibitem[{{Leroy} {et~al}\mbox{.}(2008){Leroy}, {Walter}, {Brinks}, {Bigiel},
  {de Blok}, {Madore}, \& {Thornley}}]{leroy08}
{Leroy} A.~K., {Walter} F., {Brinks} E., {Bigiel} F., {de Blok} W.~J.~G.,
  {Madore} B., {Thornley} M.~D., 2008, \aj, 136, 2782

\bibitem[{{Leroy} {et~al}\mbox{.}(2013){Leroy}, {Walter}, {Sandstrom},
  {Schruba}, {Munoz-Mateos}, {Bigiel}, {Bolatto}, {Brinks}, {de Blok}, {Meidt},
  {Rix}, {Rosolowsky}, {Schinnerer}, {Schuster}, \& {Usero}}]{leroy13}
{Leroy} A.~K. {et~al.}, 2013, \aj, 146, 19

\bibitem[{{Lin} \& {Faber}(1983)}]{lin83}
{Lin} D.~N.~C., {Faber} S.~M., 1983, \apjl, 266, L21

\bibitem[{{Magdis} {et~al}\mbox{.}(2012){Magdis}, {Daddi}, {Sargent}, {Elbaz},
  {Gobat}, {Dannerbauer}, {Feruglio}, {Tan}, {Rigopoulou}, {Charmandaris},
  {Dickinson}, {Reddy}, \& {Aussel}}]{magdis12a}
{Magdis} G.~E. {et~al.}, 2012, \apjl, 758, L9

\bibitem[{{Martin} {et~al}\mbox{.}(2014){Martin}, {Chambers}, {Collins},
  {Ibata}, {Rich}, {Bell}, {Bernard}, {Ferguson}, {Flewelling}, {Kaiser},
  {Magnier}, {Tonry}, \& {Wainscoat}}]{martin14}
{Martin} N.~F. {et~al.}, 2014, \apjl, 793, L14

\bibitem[{{Martin} {et~al}\mbox{.}(2013{\natexlab{a}}){Martin}, {Schlafly},
  {Slater}, {Bernard}, {Rix}, {Bell}, {Ferguson}, {Finkbeiner}, {Laevens},
  {Burgett}, {Chambers}, {Draper}, {Hodapp}, {Kaiser}, {Kudritzki}, {Magnier},
  {Metcalfe}, {Morgan}, {Price}, {Tonry}, {Wainscoat}, \& {Waters}}]{martin13b}
{Martin} N.~F. {et~al.}, 2013{\natexlab{a}}, \apjl, 779, L10

\bibitem[{{Martin} {et~al}\mbox{.}(2013{\natexlab{b}}){Martin}, {Slater},
  {Schlafly}, {Morganson}, {Rix}, {Bell}, {Laevens}, {Bernard}, {Ferguson},
  {Finkbeiner}, {Burgett}, {Chambers}, {Hodapp}, {Kaiser}, {Kudritzki},
  {Magnier}, {Morgan}, {Price}, {Tonry}, \& {Wainscoat}}]{martin13a}
{Martin} N.~F. {et~al.}, 2013{\natexlab{b}}, \apj, 772, 15

\bibitem[{{Mart{\'{\i}}nez-Delgado}
  {et~al}\mbox{.}(2005){Mart{\'{\i}}nez-Delgado}, {Butler}, {Rix}, {Franco},
  {Pe{\~n}arrubia}, {Alfaro}, \& {Dinescu}}]{md05}
{Mart{\'{\i}}nez-Delgado} D., {Butler} D.~J., {Rix} H.-W., {Franco} V.~I.,
  {Pe{\~n}arrubia} J., {Alfaro} E.~J., {Dinescu} D.~I., 2005, \apj, 633, 205

\bibitem[{{Mateo}(1998)}]{mateo98}
{Mateo} M.~L., 1998, \araa, 36, 435

\bibitem[{{Mateu} {et~al}\mbox{.}(2009){Mateu}, {Vivas}, {Zinn}, {Miller}, \&
  {Abad}}]{mateu09}
{Mateu} C., {Vivas} A.~K., {Zinn} R., {Miller} L.~R., {Abad} C., 2009, \aj,
  137, 4412

\bibitem[{{McConnachie}(2012)}]{mcconnachie12}
{McConnachie} A.~W., 2012, \aj, 144, 4

\bibitem[{{McConnachie} {et~al}\mbox{.}(2008){McConnachie}, {Huxor}, {Martin},
  {Irwin}, {Chapman}, {Fahlman}, {Ferguson}, {Ibata}, {Lewis}, {Richer}, \&
  {Tanvir}}]{mcconnachie08}
{McConnachie} A.~W. {et~al.}, 2008, \apj, 688, 1009

\bibitem[{{McConnachie} {et~al}\mbox{.}(2009){McConnachie}, {Irwin}, {Ibata},
  {Dubinski}, {Widrow}, {Martin}, {C{\^o}t{\'e}}, {Dotter}, {Navarro},
  {Ferguson}, {Puzia}, {Lewis}, {Babul}, {Barmby}, {Bienaym{\'e}}, {Chapman},
  {Cockcroft}, {Collins}, {Fardal}, {Harris}, {Huxor}, {Mackey},
  {Pe{\~n}arrubia}, {Rich}, {Richer}, {Siebert}, {Tanvir}, {Valls-Gabaud}, \&
  {Venn}}]{mcconnachie09}
{McConnachie} A.~W. {et~al.}, 2009, \nat, 461, 66

\bibitem[{{McGee} {et~al}\mbox{.}(2011){McGee}, {Balogh}, {Wilman}, {Bower},
  {Mulchaey}, {Parker}, \& {Oemler}}]{mcgee11}
{McGee} S.~L., {Balogh} M.~L., {Wilman} D.~J., {Bower} R.~G., {Mulchaey} J.~S.,
  {Parker} L.~C., {Oemler} A., 2011, \mnras, 413, 996

\bibitem[{{McGee}, {Bower} \& {Balogh}(2014){McGee}, {Bower}, \&
  {Balogh}}]{mcgee14}
{McGee} S.~L., {Bower} R.~G., {Balogh} M.~L., 2014, \mnras, 442, L105

\bibitem[{{Miller} \& {Bregman}(2015)}]{miller15}
{Miller} M.~J., {Bregman} J.~N., 2015, \apj, 800, 14

\bibitem[{{Mok} {et~al}\mbox{.}(2014){Mok}, {Balogh}, {McGee}, {Wilman},
  {Finoguenov}, {Tanaka}, {Bower}, {Hou}, {Mulchaey}, \& {Parker}}]{mok14}
{Mok} A. {et~al.}, 2014, \mnras, 438, 3070

\bibitem[{{Momany} {et~al}\mbox{.}(2004){Momany}, {Zaggia}, {Bonifacio},
  {Piotto}, {De Angeli}, {Bedin}, \& {Carraro}}]{momany04}
{Momany} Y., {Zaggia} S.~R., {Bonifacio} P., {Piotto} G., {De Angeli} F.,
  {Bedin} L.~R., {Carraro} G., 2004, \aap, 421, L29

\bibitem[{{Moore} {et~al}\mbox{.}(1996){Moore}, {Katz}, {Lake}, {Dressler}, \&
  {Oemler}}]{moore96}
{Moore} B., {Katz} N., {Lake} G., {Dressler} A., {Oemler} A., 1996, \nat, 379,
  613

\bibitem[{{Moster}, {Naab} \& {White}(2013){Moster}, {Naab}, \&
  {White}}]{moster13}
{Moster} B.~P., {Naab} T., {White} S.~D.~M., 2013, \mnras, 428, 3121

\bibitem[{{Muzzin} {et~al}\mbox{.}(2014){Muzzin}, {van der Burg}, {McGee},
  {Balogh}, {Franx}, {Hoekstra}, {Hudson}, {Noble}, {Taranu}, {Webb}, {Wilson},
  \& {Yee}}]{muzzin14}
{Muzzin} A. {et~al.}, 2014, \apj, 796, 65

\bibitem[{{Noeske} {et~al}\mbox{.}(2007){Noeske}, {Weiner}, {Faber},
  {Papovich}, {Koo}, {Somerville}, {Bundy}, {Conselice}, {Newman},
  {Schiminovich}, {Le Floc'h}, {Coil}, {Rieke}, {Lotz}, {Primack}, {Barmby},
  {Cooper}, {Davis}, {Ellis}, {Fazio}, {Guhathakurta}, {Huang}, {Kassin},
  {Martin}, {Phillips}, {Rich}, {Small}, {Willmer}, \& {Wilson}}]{noeske07}
{Noeske} K.~G. {et~al.}, 2007, \apjl, 660, L43

\bibitem[{{O{\~n}orbe} {et~al}\mbox{.}(2015){O{\~n}orbe}, {Boylan-Kolchin},
  {Bullock}, {Hopkins}, {Ker{\v e}s}, {Faucher-Gigu{\`e}re}, {Quataert}, \&
  {Murray}}]{onorbe15}
{O{\~n}orbe} J., {Boylan-Kolchin} M., {Bullock} J.~S., {Hopkins} P.~F., {Ker{\v
  e}s} D., {Faucher-Gigu{\`e}re} C.-A., {Quataert} E., {Murray} N., 2015, ArXiv
  e-prints

\bibitem[{{Okamoto} {et~al}\mbox{.}(2008){Okamoto}, {Arimoto}, {Yamada}, \&
  {Onodera}}]{okamoto08}
{Okamoto} S., {Arimoto} N., {Yamada} Y., {Onodera} M., 2008, \aap, 487, 103

\bibitem[{{Pasquali} {et~al}\mbox{.}(2010){Pasquali}, {Gallazzi}, {Fontanot},
  {van den Bosch}, {De Lucia}, {Mo}, \& {Yang}}]{pasquali10}
{Pasquali} A., {Gallazzi} A., {Fontanot} F., {van den Bosch} F.~C., {De Lucia}
  G., {Mo} H.~J., {Yang} X., 2010, \mnras, 407, 937

\bibitem[{{Peng} {et~al}\mbox{.}(2010){Peng}, {Lilly}, {Kova{\v c}},
  {Bolzonella}, {Pozzetti}, {Renzini}, {Zamorani}, {Ilbert}, {Knobel},
  {Iovino}, {Maier}, {Cucciati}, {Tasca}, {Carollo}, {Silverman}, {Kampczyk},
  {de Ravel}, {Sanders}, {Scoville}, {Contini}, {Mainieri}, {Scodeggio},
  {Kneib}, {Le F{\`e}vre}, {Bardelli}, {Bongiorno}, {Caputi}, {Coppa}, {de la
  Torre}, {Franzetti}, {Garilli}, {Lamareille}, {Le Borgne}, {Le Brun},
  {Mignoli}, {Perez Montero}, {Pello}, {Ricciardelli}, {Tanaka}, {Tresse},
  {Vergani}, {Welikala}, {Zucca}, {Oesch}, {Abbas}, {Barnes}, {Bordoloi},
  {Bottini}, {Cappi}, {Cassata}, {Cimatti}, {Fumana}, {Hasinger}, {Koekemoer},
  {Leauthaud}, {Maccagni}, {Marinoni}, {McCracken}, {Memeo}, {Meneux}, {Nair},
  {Porciani}, {Presotto}, \& {Scaramella}}]{peng10}
{Peng} Y.-j. {et~al.}, 2010, \apj, 721, 193

\bibitem[{P\'erez \& Granger(2007)}]{ipython}
P\'erez F., Granger B.~E., 2007, Computing in Science and Engineering, 9, 21

\bibitem[{{Phillips} {et~al}\mbox{.}(2014){Phillips}, {Wheeler},
  {Boylan-Kolchin}, {Bullock}, {Cooper}, \& {Tollerud}}]{phillips14}
{Phillips} J.~I., {Wheeler} C., {Boylan-Kolchin} M., {Bullock} J.~S., {Cooper}
  M.~C., {Tollerud} E.~J., 2014, \mnras, 437, 1930

\bibitem[{{Phillips} {et~al}\mbox{.}(2015){Phillips}, {Wheeler}, {Cooper},
  {Boylan-Kolchin}, {Bullock}, \& {Tollerud}}]{phillips15}
{Phillips} J.~I., {Wheeler} C., {Cooper} M.~C., {Boylan-Kolchin} M., {Bullock}
  J.~S., {Tollerud} E., 2015, \mnras, 447, 698

\bibitem[{{Popping}, {Behroozi} \& {Peeples}(2015){Popping}, {Behroozi}, \&
  {Peeples}}]{popping15}
{Popping} G., {Behroozi} P.~S., {Peeples} M.~S., 2015, \mnras, 449, 477

\bibitem[{{Rasmussen} {et~al}\mbox{.}(2009){Rasmussen}, {Sommer-Larsen},
  {Pedersen}, {Toft}, {Benson}, {Bower}, \& {Grove}}]{rasmussen09}
{Rasmussen} J., {Sommer-Larsen} J., {Pedersen} K., {Toft} S., {Benson} A.,
  {Bower} R.~G., {Grove} L.~F., 2009, \apj, 697, 79

\bibitem[{{Rees}(1986)}]{rees86}
{Rees} M.~J., 1986, \mnras, 218, 25P

\bibitem[{{Rocha}, {Peter} \& {Bullock}(2012){Rocha}, {Peter}, \&
  {Bullock}}]{rocha12}
{Rocha} M., {Peter} A.~H.~G., {Bullock} J., 2012, \mnras, 425, 231

\bibitem[{{Saintonge} {et~al}\mbox{.}(2011){Saintonge}, {Kauffmann}, {Wang},
  {Kramer}, {Tacconi}, {Buchbender}, {Catinella}, {Graci{\'a}-Carpio},
  {Cortese}, {Fabello}, {Fu}, {Genzel}, {Giovanelli}, {Guo}, {Haynes},
  {Heckman}, {Krumholz}, {Lemonias}, {Li}, {Moran}, {Rodriguez-Fernandez},
  {Schiminovich}, {Schuster}, \& {Sievers}}]{saintonge11}
{Saintonge} A. {et~al.}, 2011, \mnras, 415, 61

\bibitem[{{Saintonge} {et~al}\mbox{.}(2013){Saintonge}, {Lutz}, {Genzel},
  {Magnelli}, {Nordon}, {Tacconi}, {Baker}, {Bandara}, {Berta}, {F{\"o}rster
  Schreiber}, {Poglitsch}, {Sturm}, {Wuyts}, \& {Wuyts}}]{saintogne13}
{Saintonge} A. {et~al.}, 2013, \apj, 778, 2

\bibitem[{{Salim} {et~al}\mbox{.}(2007){Salim}, {Rich}, {Charlot},
  {Brinchmann}, {Johnson}, {Schiminovich}, {Seibert}, {Mallery}, {Heckman},
  {Forster}, {Friedman}, {Martin}, {Morrissey}, {Neff}, {Small}, {Wyder},
  {Bianchi}, {Donas}, {Lee}, {Madore}, {Milliard}, {Szalay}, {Welsh}, \&
  {Yi}}]{salim07}
{Salim} S. {et~al.}, 2007, \apjs, 173, 267

\bibitem[{{Sand} {et~al}\mbox{.}(2009){Sand}, {Olszewski}, {Willman},
  {Zaritsky}, {Seth}, {Harris}, {Piatek}, \& {Saha}}]{sand09}
{Sand} D.~J., {Olszewski} E.~W., {Willman} B., {Zaritsky} D., {Seth} A.,
  {Harris} J., {Piatek} S., {Saha} A., 2009, \apj, 704, 898

\bibitem[{{Sand} {et~al}\mbox{.}(2010){Sand}, {Seth}, {Olszewski}, {Willman},
  {Zaritsky}, \& {Kallivayalil}}]{sand10}
{Sand} D.~J., {Seth} A., {Olszewski} E.~W., {Willman} B., {Zaritsky} D.,
  {Kallivayalil} N., 2010, \apj, 718, 530

\bibitem[{{Schiminovich} {et~al}\mbox{.}(2010){Schiminovich}, {Catinella},
  {Kauffmann}, {Fabello}, {Wang}, {Hummels}, {Lemonias}, {Moran}, {Wu},
  {Giovanelli}, {Haynes}, {Heckman}, {Basu-Zych}, {Blanton}, {Brinchmann},
  {Budav{\'a}ri}, {Gon{\c c}alves}, {Johnson}, {Kennicutt}, {Madore}, {Martin},
  {Rich}, {Tacconi}, {Thilker}, {Wild}, \& {Wyder}}]{schiminovich10}
{Schiminovich} D. {et~al.}, 2010, \mnras, 408, 919

\bibitem[{{Skillman}, {C{\^o}t{\'e}} \& {Miller}(2003){Skillman},
  {C{\^o}t{\'e}}, \& {Miller}}]{skillman03}
{Skillman} E.~D., {C{\^o}t{\'e}} S., {Miller} B.~W., 2003, \aj, 125, 593

\bibitem[{{Slater} \& {Bell}(2014)}]{slater14}
{Slater} C.~T., {Bell} E.~F., 2014, \apj, 792, 141

\bibitem[{{Sohn} {et~al}\mbox{.}(2013){Sohn}, {Besla}, {van der Marel},
  {Boylan-Kolchin}, {Majewski}, \& {Bullock}}]{sohn13}
{Sohn} S.~T., {Besla} G., {van der Marel} R.~P., {Boylan-Kolchin} M.,
  {Majewski} S.~R., {Bullock} J.~S., 2013, \apj, 768, 139

\bibitem[{{Somerville}, {Popping} \& {Trager}(2015){Somerville}, {Popping}, \&
  {Trager}}]{somerville15}
{Somerville} R.~S., {Popping} G., {Trager} S.~C., 2015, ArXiv e-prints

\bibitem[{{Speagle} {et~al}\mbox{.}(2014){Speagle}, {Steinhardt}, {Capak}, \&
  {Silverman}}]{speagle14}
{Speagle} J.~S., {Steinhardt} C.~L., {Capak} P.~L., {Silverman} J.~D., 2014,
  \apjs, 214, 15

\bibitem[{{Spekkens} {et~al}\mbox{.}(2014){Spekkens}, {Urbancic}, {Mason},
  {Willman}, \& {Aguirre}}]{spekkens14}
{Spekkens} K., {Urbancic} N., {Mason} B.~S., {Willman} B., {Aguirre} J.~E.,
  2014, \apjl, 795, L5

\bibitem[{{Stewart} {et~al}\mbox{.}(2009){Stewart}, {Bullock}, {Barton}, \&
  {Wechsler}}]{stewart09}
{Stewart} K.~R., {Bullock} J.~S., {Barton} E.~J., {Wechsler} R.~H., 2009, \apj,
  702, 1005

\bibitem[{{Strateva} {et~al}\mbox{.}(2001){Strateva}, {Ivezi{\'c}}, {Knapp},
  {Narayanan}, {Strauss}, {Gunn}, {Lupton}, {Schlegel}, {Bahcall}, {Brinkmann},
  {Brunner}, {Budav{\'a}ri}, {Csabai}, {Castander}, {Doi}, {Fukugita}, {Gy{\H
  o}ry}, {Hamabe}, {Hennessy}, {Ichikawa}, {Kunszt}, {Lamb}, {McKay},
  {Okamura}, {Racusin}, {Sekiguchi}, {Schneider}, {Shimasaku}, \&
  {York}}]{strateva01}
{Strateva} I. {et~al.}, 2001, \aj, 122, 1861

\bibitem[{{Sun} {et~al}\mbox{.}(2007){Sun}, {Jones}, {Forman}, {Vikhlinin},
  {Donahue}, \& {Voit}}]{sun07}
{Sun} M., {Jones} C., {Forman} W., {Vikhlinin} A., {Donahue} M., {Voit} M.,
  2007, \apj, 657, 197

\bibitem[{{Tacconi} {et~al}\mbox{.}(2010){Tacconi}, {Genzel}, {Neri}, {Cox},
  {Cooper}, {Shapiro}, {Bolatto}, {Bouch{\'e}}, {Bournaud}, {Burkert},
  {Combes}, {Comerford}, {Davis}, {Schreiber}, {Garcia-Burillo},
  {Gracia-Carpio}, {Lutz}, {Naab}, {Omont}, {Shapley}, {Sternberg}, \&
  {Weiner}}]{tacconi10}
{Tacconi} L.~J. {et~al.}, 2010, \nat, 463, 781

\bibitem[{{Tacconi} {et~al}\mbox{.}(2013){Tacconi}, {Neri}, {Genzel}, {Combes},
  {Bolatto}, {Cooper}, {Wuyts}, {Bournaud}, {Burkert}, {Comerford}, {Cox},
  {Davis}, {F{\"o}rster Schreiber}, {Garc{\'{\i}}a-Burillo}, {Gracia-Carpio},
  {Lutz}, {Naab}, {Newman}, {Omont}, {Saintonge}, {Shapiro Griffin}, {Shapley},
  {Sternberg}, \& {Weiner}}]{tacconi13}
{Tacconi} L.~J. {et~al.}, 2013, \apj, 768, 74

\bibitem[{{Tollerud} {et~al}\mbox{.}(2011){Tollerud}, {Boylan-Kolchin},
  {Barton}, {Bullock}, \& {Trinh}}]{tollerud11}
{Tollerud} E.~J., {Boylan-Kolchin} M., {Barton} E.~J., {Bullock} J.~S., {Trinh}
  C.~Q., 2011, \apj, 738, 102

\bibitem[{{Tollerud}, {Boylan-Kolchin} \& {Bullock}(2014){Tollerud},
  {Boylan-Kolchin}, \& {Bullock}}]{tollerud14}
{Tollerud} E.~J., {Boylan-Kolchin} M., {Bullock} J.~S., 2014, \mnras, 440, 3511

\bibitem[{{Tollerud} {et~al}\mbox{.}(2008){Tollerud}, {Bullock}, {Strigari}, \&
  {Willman}}]{tollerud08}
{Tollerud} E.~J., {Bullock} J.~S., {Strigari} L.~E., {Willman} B., 2008, \apj,
  688, 277

\bibitem[{{van den Bosch} {et~al}\mbox{.}(2008){van den Bosch}, {Aquino},
  {Yang}, {Mo}, {Pasquali}, {McIntosh}, {Weinmann}, \& {Kang}}]{vdb08}
{van den Bosch} F.~C., {Aquino} D., {Yang} X., {Mo} H.~J., {Pasquali} A.,
  {McIntosh} D.~H., {Weinmann} S.~M., {Kang} X., 2008, \mnras, 387, 79

\bibitem[{van~der Walt, Colbert \& Varoquaux(2011)van~der Walt, Colbert, \&
  Varoquaux}]{numpy}
van~der Walt S., Colbert S.~C., Varoquaux G., 2011, CoRR, abs/1102.1523

\bibitem[{{Wang} {et~al}\mbox{.}(2014){Wang}, {Sales}, {Henriques}, \&
  {White}}]{wang14}
{Wang} W., {Sales} L.~V., {Henriques} B.~M.~B., {White} S.~D.~M., 2014, \mnras,
  442, 1363

\bibitem[{{Weinmann} {et~al}\mbox{.}(2010){Weinmann}, {Kauffmann}, {von der
  Linden}, \& {De Lucia}}]{weinmann10}
{Weinmann} S.~M., {Kauffmann} G., {von der Linden} A., {De Lucia} G., 2010,
  \mnras, 406, 2249

\bibitem[{{Weinmann} {et~al}\mbox{.}(2011){Weinmann}, {Lisker}, {Guo}, {Meyer},
  \& {Janz}}]{weinmann11}
{Weinmann} S.~M., {Lisker} T., {Guo} Q., {Meyer} H.~T., {Janz} J., 2011,
  \mnras, 416, 1197

\bibitem[{{Weinmann} {et~al}\mbox{.}(2012){Weinmann}, {Pasquali},
  {Oppenheimer}, {Finlator}, {Mendel}, {Crain}, \& {Macci{\`o}}}]{weinmann12}
{Weinmann} S.~M., {Pasquali} A., {Oppenheimer} B.~D., {Finlator} K., {Mendel}
  J.~T., {Crain} R.~A., {Macci{\`o}} A.~V., 2012, \mnras, 426, 2797

\bibitem[{{Weisz} {et~al}\mbox{.}(2014{\natexlab{a}}){Weisz}, {Dolphin},
  {Skillman}, {Holtzman}, {Gilbert}, {Dalcanton}, \& {Williams}}]{weisz14a}
{Weisz} D.~R., {Dolphin} A.~E., {Skillman} E.~D., {Holtzman} J., {Gilbert}
  K.~M., {Dalcanton} J.~J., {Williams} B.~F., 2014{\natexlab{a}}, \apj, 789,
  147

\bibitem[{{Weisz} {et~al}\mbox{.}(2014{\natexlab{b}}){Weisz}, {Dolphin},
  {Skillman}, {Holtzman}, {Gilbert}, {Dalcanton}, \& {Williams}}]{weisz14b}
{Weisz} D.~R., {Dolphin} A.~E., {Skillman} E.~D., {Holtzman} J., {Gilbert}
  K.~M., {Dalcanton} J.~J., {Williams} B.~F., 2014{\natexlab{b}}, \apj, 789,
  148

\bibitem[{{Weisz} {et~al}\mbox{.}(2015){Weisz}, {Dolphin}, {Skillman},
  {Holtzman}, {Gilbert}, {Dalcanton}, \& {Williams}}]{weisz15}
{Weisz} D.~R., {Dolphin} A.~E., {Skillman} E.~D., {Holtzman} J., {Gilbert}
  K.~M., {Dalcanton} J.~J., {Williams} B.~F., 2015, \apj, 804, 136

\bibitem[{{Wetzel}, {Deason} \& {Garrison-Kimmel}(2015){Wetzel}, {Deason}, \&
  {Garrison-Kimmel}}]{wetzel15}
{Wetzel} A.~R., {Deason} A.~J., {Garrison-Kimmel} S., 2015, \apj, 807, 49

\bibitem[{{Wetzel} {et~al}\mbox{.}(2013){Wetzel}, {Tinker}, {Conroy}, \& {van
  den Bosch}}]{wetzel13}
{Wetzel} A.~R., {Tinker} J.~L., {Conroy} C., {van den Bosch} F.~C., 2013,
  \mnras, 432, 336

\bibitem[{{Wetzel}, {Tollerud} \& {Weisz}(2015){Wetzel}, {Tollerud}, \&
  {Weisz}}]{wetzel15b}
{Wetzel} A.~R., {Tollerud} E.~J., {Weisz} D.~R., 2015, \apjl, 808, L27

\bibitem[{{Wheeler} {et~al}\mbox{.}(2015){Wheeler}, {Onorbe}, {Bullock},
  {Boylan-Kolchin}, {Elbert}, {Garrison-Kimmel}, {Hopkins}, \&
  {Keres}}]{wheeler15}
{Wheeler} C., {Onorbe} J., {Bullock} J.~S., {Boylan-Kolchin} M., {Elbert}
  O.~D., {Garrison-Kimmel} S., {Hopkins} P.~F., {Keres} D., 2015, ArXiv
  e-prints

\bibitem[{{Wheeler} {et~al}\mbox{.}(2014){Wheeler}, {Phillips}, {Cooper},
  {Boylan-Kolchin}, \& {Bullock}}]{wheeler14}
{Wheeler} C., {Phillips} J.~I., {Cooper} M.~C., {Boylan-Kolchin} M., {Bullock}
  J.~S., 2014, \mnras, 442, 1396

\bibitem[{{Woo} {et~al}\mbox{.}(2013){Woo}, {Dekel}, {Faber}, {Noeske}, {Koo},
  {Gerke}, {Cooper}, {Salim}, {Dutton}, {Newman}, {Weiner}, {Bundy}, {Willmer},
  {Davis}, \& {Yan}}]{woo13}
{Woo} J. {et~al.}, 2013, \mnras, 428, 3306

\bibitem[{{Wyithe} \& {Loeb}(2006)}]{wyithe06}
{Wyithe} J.~S.~B., {Loeb} A., 2006, \nat, 441, 322

\bibitem[{{Yang} {et~al}\mbox{.}(2007){Yang}, {Mo}, {van den Bosch},
  {Pasquali}, {Li}, \& {Barden}}]{yang07}
{Yang} X., {Mo} H.~J., {van den Bosch} F.~C., {Pasquali} A., {Li} C., {Barden}
  M., 2007, \apj, 671, 153

\bibitem[{{York} {et~al}\mbox{.}(2000){York}, {Adelman}, {Anderson},
  {Anderson}, {Annis}, {Bahcall}, {Bakken}, {Barkhouser}, {Bastian}, {Berman},
  {Boroski}, {Bracker}, {Briegel}, {Briggs}, {Brinkmann}, {Brunner}, {Burles},
  {Carey}, {Carr}, {Castander}, {Chen}, {Colestock}, {Connolly}, {Crocker},
  {Csabai}, {Czarapata}, {Davis}, {Doi}, {Dombeck}, {Eisenstein}, {Ellman},
  {Elms}, {Evans}, {Fan}, {Federwitz}, {Fiscelli}, {Friedman}, {Frieman},
  {Fukugita}, {Gillespie}, {Gunn}, {Gurbani}, {de Haas}, {Haldeman}, {Harris},
  {Hayes}, {Heckman}, {Hennessy}, {Hindsley}, {Holm}, {Holmgren}, {Huang},
  {Hull}, {Husby}, {Ichikawa}, {Ichikawa}, {Ivezi{\'c}}, {Kent}, {Kim},
  {Kinney}, {Klaene}, {Kleinman}, {Kleinman}, {Knapp}, {Korienek}, {Kron},
  {Kunszt}, {Lamb}, {Lee}, {Leger}, {Limmongkol}, {Lindenmeyer}, {Long},
  {Loomis}, {Loveday}, {Lucinio}, {Lupton}, {MacKinnon}, {Mannery}, {Mantsch},
  {Margon}, {McGehee}, {McKay}, {Meiksin}, {Merelli}, {Monet}, {Munn},
  {Narayanan}, {Nash}, {Neilsen}, {Neswold}, {Newberg}, {Nichol}, {Nicinski},
  {Nonino}, {Okada}, {Okamura}, {Ostriker}, {Owen}, {Pauls}, {Peoples},
  {Peterson}, {Petravick}, {Pier}, {Pope}, {Pordes}, {Prosapio},
  {Rechenmacher}, {Quinn}, {Richards}, {Richmond}, {Rivetta}, {Rockosi},
  {Ruthmansdorfer}, {Sandford}, {Schlegel}, {Schneider}, {Sekiguchi}, {Sergey},
  {Shimasaku}, {Siegmund}, {Smee}, {Smith}, {Snedden}, {Stone}, {Stoughton},
  {Strauss}, {Stubbs}, {SubbaRao}, {Szalay}, {Szapudi}, {Szokoly}, {Thakar},
  {Tremonti}, {Tucker}, {Uomoto}, {Vanden Berk}, {Vogeley}, {Waddell}, {Wang},
  {Watanabe}, {Weinberg}, {Yanny}, {Yasuda}, \& {SDSS Collaboration}}]{york00}
{York} D.~G. {et~al.}, 2000, \aj, 120, 1579

\end{thebibliography}

\label{lastpage}
\end{document}